\documentclass[11pt,a4paper]{report}

\usepackage[left=35mm, top=30mm, bottom=30mm, right=35mm]{geometry}
 
\usepackage{graphicx} 
\usepackage{epstopdf} 
\usepackage{multicol} 
\usepackage{calc}     
\usepackage{enumitem} 
\usepackage{titlesec} 
\usepackage{lipsum}   
\usepackage{fancyhdr} 
\usepackage{pdfpages} 
\usepackage{tikz}     
\usepackage{fourier-orns} 
\usepackage{textcomp} 
\usepackage{mathtools}
\usepackage{algorithm,algorithmic}
\usepackage{booktabs} 
\usepackage{multirow} 
\usepackage{caption} 
\usepackage{chngcntr}

\graphicspath{ {../img/} }
\titleformat{\chapter}{\normalfont\huge}{\thechapter}{20pt}{\huge\it}
\newlength\myindent
\newcommand\bindent{%
  \begingroup
  \setlength{\itemindent}{\myindent}
  \addtolength{\algorithmicindent}{\myindent}
}
\newcommand\eindent{\endgroup}
\counterwithout{figure}{chapter}
\counterwithout{table}{chapter}
\newcommand\MyHead[2]{%
  \multicolumn{1}{l}{\parbox{#1}{\flushleft #2}}
}

\begin{document}

\newpage
\thispagestyle{empty}
\begin{titlepage}
	\centering
	\large{Department of Medical Biophysics and Biochemistry \\
	Karolinska Institutet, Stockholm, Sweden \par}
	\vspace{1cm}
	\vspace{1cm}
	{\LARGE\bfseries eQTL Mapping and Inherited Risk Enrichment Analysis - A Systems Biology Approach for Coronary Artery Disease\par}
	\vspace{2cm}
	{\Large Hassan Foroughi Asl\par}
	\vspace{4cm}
	\includegraphics[scale=0.5]{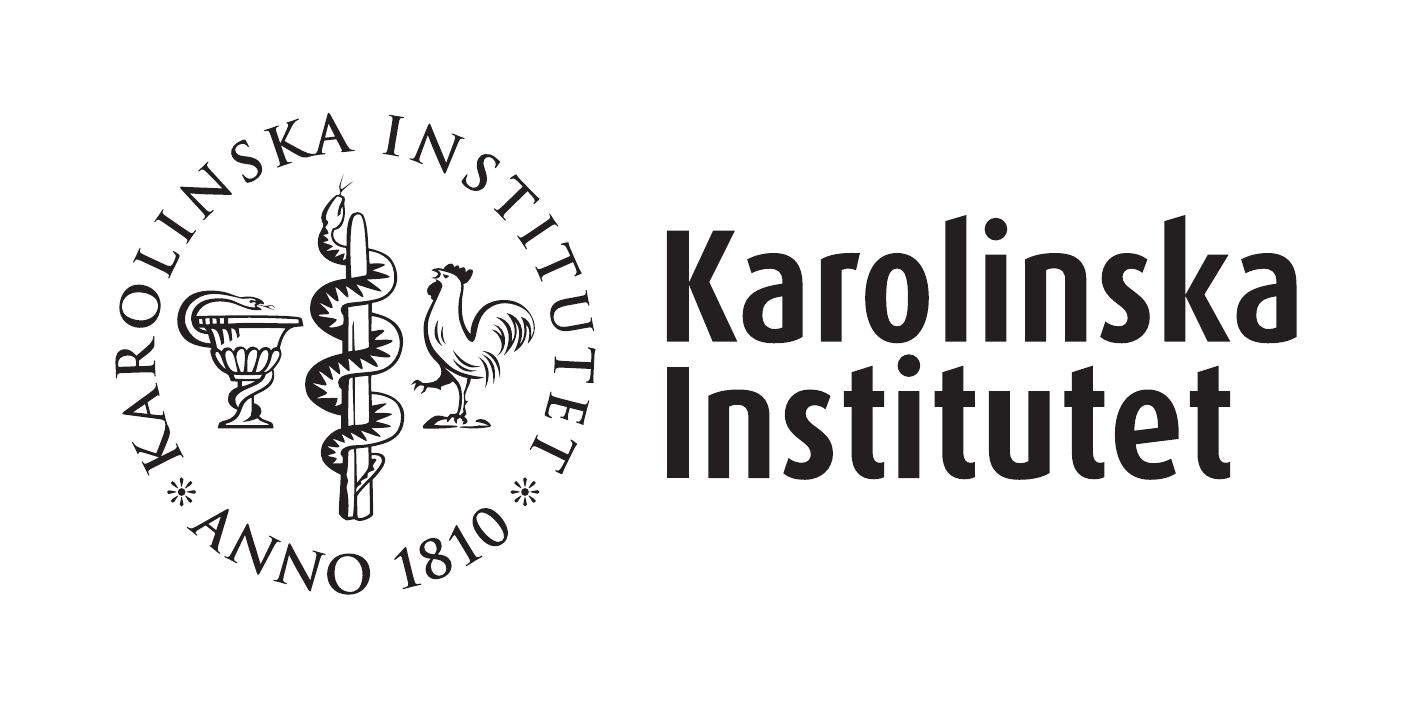}
	\vspace{1cm}
	\center{Stockholm 2016}
\end{titlepage}

\newpage
\thispagestyle{empty}
\begin{raggedright}
{~\par}
\vspace{15cm}
{All previously published papers were reproduced with permission from the publisher. \par}
\vspace{1cm}
{Published by Karolinska Institutet. Printed by Eprint AB 2016. \par}
\vspace{1cm}
{\copyright ~Hassan Foroughi Asl, 2016\par}
\vspace{1cm}
{\indent
{~~~~ISBN 978-91-7676-251-6}}
\end{raggedright}

\newpage
\thispagestyle{empty}
{
	\raggedright
	{ \uppercase{eQTL Mapping and Inherited Risk Enrichment Analysis - A Systems Biology Approach for Coronary Artery Disease}\par}
	\vspace{0.25cm}
	{\uppercase {thesis for doctoral degree (Ph.D.)}\par}
	\vspace{0.25cm}
	{By \par}
	\vspace{0.25cm}
	{\Large\textbf{Hassan Foroughi Asl}\par}
	\vspace{1cm}
	{\textit{Principal Supervisor:}\\
	\textbf{Johan Bj\"orkegren}\\
	Karolinska Institutet, Department of Medical Biochemistry and Biophysics\par}
	\vspace{0.5cm}
	{\textit{Co-Supervisor(s):}\\
	\textbf{Josefin Skogsberg}\\
	Karolinska Institutet, Department of Medical Biochemistry and Biophysics\par}
	\vspace{0.5cm}
	{\textbf{Tom Michoel}\\
	University of Edinburgh, The Roslin Institute, Division of Genetics and Genomics\par}
	\vspace{0.5cm}
	{\textbf{Christer Betsholtz}\\
	Uppsala Universitet, Department of Immunology, Genetics, and Pathology\par}
	\vspace{0.5cm}
	{\textit{Opponent:}\par
	\textbf{Assoc. Prof. Carmen Argmann}\\
	Icahn School of Medicine at Mount Sinai, Department of Genetics and Genomics\par}
	\vspace{0.5cm}
	{\textit{Examination board:}\par}
	{\textbf{Assoc. Prof. Leonid Padyukov}\\
	Karolinska Institutet, Department of Medicine (Solna)\par}
	\vspace{0.5cm}
	{\textbf{Prof. Lars Lind}\\
	Uppsala Universitet, Department of Medical Sciences\par}
	\vspace{0.5cm}
	{\textbf{Assoc. Prof. Lars Feuk}\\
	Uppsala Universitet, Department of Immunology, Genetics, and Pathology}
}
\clearpage

\newpage
\thispagestyle{empty}
{~\par}
\clearpage

\newpage
\thispagestyle{empty}
\begin{flushright}
{~\par}
\vspace{18cm}
\textit{{``Approximation of the truth is all that we have"} \\
Alan Chalmers}
\end{flushright}
\clearpage

\thispagestyle{empty}
{~\par}
\clearpage

\newpage
\thispagestyle{empty}
\begin{center}
\Large\textbf{Abstract}
\end{center}
{
{Despite extensive research during the last decades, coronary artery disease (CAD) remains the number one cause of death, responsible for near 50\% of global mortality. A main reason for this is that CAD has a complex inheritance and etiology that unlike rare single gene disorders cannot fully be understood from studies of of genes one-by-one.In parallel, studies that simultaneously assess multiple, functionally associated genes are warranted. For this reason we undertook the Stockholm Atherosclerosis Gene Expression (STAGE) study that besides careful clinical characterization and genome-wide DNA genotyping also assessed the global gene expression profiles from seven CAD-relevant vascular and metabolic tissues.}\newline \indent
{In paper I, we used STAGE to develop a bioinformatics tool for efficient eQTL mapping called kruX based on Kruskal-Wallis statistics test. kruX excels in detecting a higher proportion of nonlinear expression quantitative expression traits loci (eQTLs) compared to other established methods. This tool was developed for Python, MATLAB, and R and is available online. In paper II, we applied kruX to detect eQTLs across the seven tissues in STAGE and assessed their tissue specificity. A tool for analyzing inherited risk enrichment was also developed assessing CAD association (i.e., risk enrichment) of STAGE eQTLs according to genome-wide association studies (GWAS) of CAD. We found that eQTLs active across multiple vascular and metabolic tissues are more enriched in inherited risk for CAD than tissue-specific eQTLs. In paper III, we integrate the analysis of STAGE data with data from GWAS of CAD to identify 30 regulatory-gene networks causal for CAD. In paper IV, we again used kruX to investigate STAGE eQTLs for three established candidate genes in CAD and atherosclerosis (ALOX5, ALOX5AP, and LTA4H). In addition, we used the Athero-Express dataset of genotype and atherosclerotic carotid plaque characteristics to further elucidate the role of these genes in atherosclerosis development.}\newline \indent
{In sum, in this thesis report we show that by integrating GWAS with genetics of gene expression studies like STAGE, we can advance our understanding from the perspective of multiple genes and gene variants acting in conjunction to cause CAD in the form of regulatory gene networks. This is done through developing new bioinformatics tools and applying them to disease-specific, genetics of global gene expression studies like STAGE. These tools are necessary to go beyond our current limited single-gene understanding of complex traits, like CAD.}
}
\clearpage

\newpage
\thispagestyle{empty}
\begin{center}
{\Large\textbf{List of Publications}\\}
\end{center}
\begin{flushleft}
{This thesis is based on the publications listed below.}
\end{flushleft}

\begin{description}[labelwidth=\widthof{IV}]
\item[\textit{I}]{
{Jianlong Qi, \textbf{Hassan Foroughi Asl}, Johan Bj\"orkegren, Tom Michoel.}
\textbf{kruX: matrix-based non-parametric eQTL discovery.}
\textit{BMC Bioinformatics 2014 Jan 14; 15:11.doi: 10.1186/1471-2105-15-11} 
}
\item[\textit{II}]{
{\textbf{Hassan Foroughi Asl}, Husain A Talukdar, Alida SD Kindt, Rajeev K Jain, Raili Ermel, Arno Ruusalepp, Khanh-Dung H Nguyen, Radu Dobrin, Dermot F Reilly, Heribert Schunkert, Nilesh J Samani, Ingrid Br{\ae}nne, Jeanette Erdmann, Olle Melander, Jianlong Qi, Torbj\"orn Ivert, Josefin Skogsberg, Eric E Schadt, Tom Michoel, Johan L.M. Bj\"orkegren.}
\textbf{Expression quantitative trait Loci acting across multiple tissues are enriched in inherited risk for coronary artery disease.}
\textit{Circulation Cardiovascular Genetics. 2015 Apr; 8(2):305-15. doi: 10.1161/\-CIRCGENETICS.114.000640}
}
\item[\textit{III}]{
{Husain A. Talukdar, \textbf{Hassan Foroughi Asl}, Rajeev K. Jain, Raili Ermel, Arno Ruusalepp, Oscar Franz\'{e}n, Brian A. Kidd, Ben Readhead, Chiara Giannarelli, Jason C. Kovacic, Torbj\"orn Ivert, Joel T. Dudley, Mete Civelek, Aldons J. Lusis, Eric E. Schadt, Josefin Skogsberg, Tom Michoel,and Johan L.M. Bj\"orkegren.}
\textbf{Cross-Tissue Regulatory Gene Networks in Coronary Artery Disease}
\textit{Cell Systems. 2016 March; 2(3):196–-208. doi: 10.1016/\-j.cels.2016.02.002.}
}
\item[\textit{IV}]{
{Sander W van der Laan*, \textbf{Hassan Foroughi Asl}*, Pleunie van den Borne, Jessica van Setten, ME Madeleine van der Perk, Sander M van de Weg, Arjan H Schoneveld, Dominique PV de Kleijn, Tom Michoel, Johan L.M. Bj\"orkegren, Hester M den Ruijter, Folkert W Asselbergs, Paul IW de Bakker, Gerard Pasterkamp}
\textbf{Variants in ALOX5, ALOX5AP and LTA4H are not associated with atherosclerotic plaque phenotypes: the Athero Express Genomics Study.}
\textit{Atherosclerosis. 2015 Apr;239(2):528-38. doi: 10.1016/j. atherosclerosis.\\ 2015.01.018.}\\
\tiny{*Equal contribution}
}
\end{description}
\clearpage

\newpage
\thispagestyle{empty}
\begin{center}
{\Large\textbf{Other Publications}\\}
\end{center}
\begin{flushleft}
{The list below are other publications that I've been participated in during my Ph.D. studies but not included in this thesis}
\end{flushleft}

\begin{itemize}
\item{
{Ming-Mei Shang, Husain A. Talukdar, Jennifer J. Hofmann, Colin Niaudet, \textbf{Hassan Foroughi Asl}, Rajeev K. Jain, Aranzazu Rossignoli, Cecilia Cedergren, Angela Silveira, et al.}
\textbf{Lim Domain Binding 2: A Key Driver of Transendothelial Migration of Leukocytes and Atherosclerosis}
\textit{Arterioscler Thromb Vasc Biol. 2014 Sep; 34(9):2068-77. doi: 10.1161/ ATVBAHA.113.302709.}
}

\item{
{Johan L. M. Bj\"orkegren, Sara H\"agg, Husain A. Talukdar, \textbf{Hassan Foroughi Asl} , Rajeev K. Jain, Cecilia Cedergren, Ming-Mei Shang, Ar\'{a}nzazu Rossignoli, Rabbe Takolander, et al.}
\textbf{Plasma cholesterol-induced lesion networks activated before regression of early, mature, and advanced atherosclerosis}
\textit{PLoS Genet. 2014 Feb 27;10(2): e1004201. doi: 10.1371/ journal.pgen.1004201.}
}

\item{
{Ingrid Br{\ae}nne, Mete Civelek, Baiba Vilne, Antonio Di Narzo, Andrew D. Johnson, Yuqi Zhao, Benedikt Reiz, Veronica Codoni, Thomas R. Webb, \textbf{Hassan Foroughi Asl}, Stephen E. Hamby, et al.}
\textbf{Prediction of Causal Candidate Genes in Coronary Artery Disease Loci}
\textit{Arteriosclerosis, Thrombosis, and Vascular Biology. 2015 Oct;35(10):2207-17. doi: 10.1161/ATVBAH A.115.306108.}
}
\item{
{Gerard Pasterkamp, Sander W. van der Laan, Saskia Haitjema, \textbf{Hassan Foroughi Asl}, Marten A. Siemelink, Tim Bezemer, Jessica van Setten, et al.}
\textbf{Human Validation of Genes Associated With a Murine Atheros clerotic Phenotype}
\textit{Arteriosclerosis, Thrombosis, and Vascular Biology. 2016 Apr 14. doi: 10.1161/ ATVBAHA.115. 306958.}
}
\end{itemize}
\clearpage

\newpage
\thispagestyle{empty}
\pagenumbering{gobble}
\chapter*{Abbreviations}
{
\begin{tabular}{ll}
\textbf{bp}   & base pair                           \\
\textbf{CAD}  & Coronary Artery Disease             \\
\textbf{CCD}  & Common Complex Disease              \\
\textbf{CNV}  & Copy Number Variation               \\
\textbf{CVD}  & Cardiovascular Disease              \\
\textbf{DNA}  & Deoxyribonucleic Acid               \\
\textbf{EC}   & Endothelial Cell                    \\
\textbf{ED}   & Endothelial Dysfunction             \\
\textbf{eQTL} & Expression Quantitative Trait Loci  \\
\textbf{FDR}  & False Discovery Rate                \\
\textbf{GGES} & Genetics of Gene Expression Studies \\
\textbf{GWAS} & Genome-wide Association Studies     \\
\textbf{IBD}  & Identity By Descent                 \\
\textbf{IBS}  & Identical By State                  \\
\textbf{LD}   & Linkage Disequilibrium              \\
\textbf{LDL}  & Low-density Lipoprotein             \\
\textbf{MAF}  & Minor Allele Frequency              \\
\textbf{MI}   & Myocardial Infarction               \\
\textbf{mRNA} & Messenger Ribonucleic Acid          \\
\textbf{SMC}  & Smooth Muscle Cell                  \\
\textbf{SNP}  & Single-nucleotide polymorphism      \\
\textbf{TSE}  & Transcription End Site              \\
\textbf{TSS}  & Transcription Start Site            \\
\textbf{GO}   & Gene Ontology                      
\end{tabular}
}

\newpage
\thispagestyle{empty}
{~\par}
\clearpage

\newpage
\pagenumbering{gobble}
\thispagestyle{empty}
\tableofcontents
\thispagestyle{empty}
\cleardoublepage

\newpage
\thispagestyle{empty}
{~\par}
\clearpage
\newpage
\pagenumbering{arabic}
\setcounter{page}{1}
\setcounter{figure}{0}

\chapter{Background}{Over the last few decades, socioeconomic changes improved standards of living (health, hygiene, and nutritional habits), and medical advances have led to declines in mortality rates from infectious diseases and nutritional deficiencies and a shift toward degenerative diseases, such as cardiovascular disease (CVD, a disease of heart and peripheral vessels) and diabetes  \cite{Naghavi2015,Yusuf2001a,Yusuf2001b}. This so-called epidemiological transition reflects the modernization of societies, mostly in middle-income and developed countries  \cite{Yusuf2001a}. The increase in CVD-related incidents suggests that the effects of environmental and behavioral factors have been underestimated. According to the World Health Organization (WHO), CVD is the predominant cause of mortality worldwide, accounting for approximately 30\% of deaths (17.3 million people in 2013) to a different extent in men and women  \cite{Stevens2009}. Approximately 40\% of the CVD-related mortality was from coronary artery disease (CAD) and stroke  \cite{Stevens2009}.}
\section{Coronary Artery Disease: a Disease of Arteries of the Heart}
{Coronary artery disease (CAD) is caused by atherosclerosis of arteries in the heart. Also known as atherosclerotic CVD, CAD is a progressive lifelong inflammatory disease  \cite{Lusis2000,Lusis2012,Hansson2005}, whose course can be influenced by metabolic disturbances in the body  \cite{Tegner2007}. This disease mainly affects large and intermediate-sized arteries. It is characterized by gradual accumulation of inflammatory cells (leukocytes), a growing lipid core, and migration of smooth muscle cells (SMC) into the arterial wall  \cite{Lusis2000,Lusis2012}. Eventually, fatty streaks develop and mature into atherosclerotic plaques  \cite{Lusis2000,Lusis2012}.}
\subsection{Arteries and their Structure}
{Arteries are the vessels that carry blood from the heart to other organs in the body. Large arteries have three distinct layers: intima, media, and adventitia. Intima, the closest layer to the lumen, is composed of a single layer of endothelial cells (EC) and a basal membrane that attaches EC to connective tissues. The EC layer is a selectively permeable barrier for materials to pass through. It also provides strength and flexibility to the intima. Media, the middle and generally thickest layer, is mainly composed of SMC and some connective tissue. SMC are bound to the other two layers by collagen fibers. The adventitia, the outermost layer, can be thicker than media in large arteries and is mainly composed of collagen fibers, connective tissue, and perivascular nerves.}
\subsection{Molecular Events of Atherosclerosis}
{Two main physiological functions of the endothelial monolayer are to provide a nonadhesive barrier for blood to flow smoothly and to modulate constriction of SMC to dilate or constrict the vascular wall based on demand  \cite{Lusis2000}. Any change or damage to the endothelial layer that inhibits or alters its function is referred to as “endothelial dysfunction” (ED). ED can be caused by systemic factors, such as increases in blood cholesterol levels (hypercholesterolemia) or blood pressure (hypertension), or local factors, such as blood flow dynamics and inflammation  \cite{Lusis2000}. ED can alter endothelial permeability (due to shear blood flow stress) and increase the expression of adhesion molecules  \cite{Lusis2000}, setting the stage for a key initiating event in the development of atherosclerotic plaques-infiltration of low-density lipoproteins (LDL) into the subendothelial layer. The rates of infiltration and retention are accelerated by hypercholesterolemia and other systemic changes. Within the subendothelial layer, retained LDL are modified into oxidized LDL (oxLDL) and activate EC to express adhesion molecules. Circulating inflammatory cells (mainly monocytes) attach to these molecules and migrate to the subendothelium, where the monocytes differentiate into macrophages, which take up oxLDL and transform into foam-like cells, which aggregate into fatty streaks - the starting point of atherosclerotic plaque. Foam cells at the heart of the plaque may die \cite{Seimon2009} and burst in the ever-growing core. During this initial inflammatory response and foam-cell formation, SMC migrate from the media. A SMC-derived extracellular matrix forms and encapsulates the growing core. Eventually, a fibrous cap forms. Migrated inflammatory cells also secrete cell growth and proliferation factors and inflammatory cytokines (signaling proteins) that elicit an additional inflammatory response and cause SMC proliferation. In complex plaques, the growing core may become calcified  \cite{Lusis2000}.}
\subsection{Clinical Manifestation of Atherosclerosis}
{Atherosclerotic plaques can become so large that they bulge toward the vessel lumen. The plaques have two general forms (or a mixture of both): stenotic and nonstenotic  \cite{Libby2005}. Stenosis refers to narrowing of lumen, which can reduce or block blood flow. Stenotic plaques have smaller lipid core and a thick fibrous cap, whereas nonstenotic plaques have larger lipid core and a thin fibrous cap. Unlike stenotic plaques, nonstenotic plaques are vulnerable to rupture at the shoulder region from causes including physical disruption, extracellular matrix degradation by macrophages, and reduction of matrix formation by SMCs. When the plaque ruptures, a blood clot (thrombus) forms at the site of the rupture and can block blood flow in the lumen (thrombosis). The thrombus can break loose from the vessel wall, becoming an embolus that travels with the circulation. Thrombosis and emboli can lead to complications such as myocardial infarction and stroke.}
\subsection{CAD is a Multifactorial Disease}
{CAD is a common disease whose development is affected by many risk factors. For that reason it is referred to as common complex disease (CCD). The risk factors range from isolated local molecular events at the atherosclerotic plaque site or distal molecular events in metabolic tissues such as the liver. These risk factors can be either modifiable or nonmodifiable and can have both direct effects (casual) and indirect effects on disease development.
The modifiable risk factors include obesity, diabetes, dyslipidemia, hypertension, inflammation, smoking and tobacco consumption, high-fat diet, and sedentary lifestyle. Examples of nonmodifiable risk factors are age, gender, and family history. Both group of risk factors are governed by genetics (e.g., family history) and its interaction with lifestyle and environmental variables (e.g., smoking and high-fat diet). Some of the major risk factors are briefly discussed below.
\begin{itemize}
\item{\textit{Family history and inheritance:} The Framingham Heart Study showed that family history is a risk factor for CAD  \cite{Mayer2007}, that there is an inheritance pattern for CAD  \cite{Wang2003,Mani2007}, the effect of mutation in single gene leading to atherosclerosis and CAD  \cite{Mani2007,Weng2005} and that it aggregates in families, reflecting the heritability of CAD  \cite{Marenberg1994,Mayer2007}.}
\item{\textit{Obesity and type 2 diabetes:} Insulin resistance is the link between obesity and CAD  \cite{Reaven2011}. Insulin resistance has been linked to waist circumference (WC; i.e., abdominal obesity)  \cite{Reaven2011}, waist-to-hip ratio (WHR)  \cite{DeKoning2007}, and body mass index (BMI). This implies that there is a direct link between obesity and CAD development risk. Therefore, body fat distribution (visceral fat and subcutaneous fat) can be predictors of CAD risk  \cite{Reaven2011}. Thus, the importance of studying molecular events within visceral fat and subcutaneous fat. Also, insulin resistence manifests itself in skeletal muscle and fat tissues, two insulin senstive tissues, as impairment in glucose transport \cite{Sowers2001}.}
\item{\textit{Dyslipidemia:} The liver has an important role in lipid metabolism, which is associated with risk for CAD. High levels of LDL, low levels of high-density lipoproteins (HDL), and high levels of triglyceride (TG) are strongly associated with increased CAD risk  \cite{Poss2011}.}
\item{\textit{Hypertension:} Hypertension is also associated with CAD risk, and elevated systolic and diastolic blood pressure, each have been linked to CAD mortality  \cite{Sowers2001}. Insulin resistence is also link to the blood flow by impairing vascular relaxation in skeletal muscle and fat tissues \cite{Sowers2001}. An important effect of hypertension is mechanical pressure and turbulent blood flow that stress the arterial wall, especially at branching points, which can contribute to disease development.}
\item{\textit{Inflammation:} Inflammation - the interaction between metabolic factors and immune mechanisms - is a key risk factor for CAD that contributes to the initiation and progression of atherosclerosis \cite{Hansson2005,Pearson2003,Libby2002}.}
\item{\textit{Smoking:} Smoking is responsible for least 10\% of all CVD incidents  \cite{Stevens2009}. Cigarette smoking in particular contributes to disease risk by increasing the risk of type 2 diabetes - itself is a major risk factor for CAD. Oxidants are thought to be the main driver of CAD risk from smoking, as almost all published research articles found no direct effect of nicotine and tobacco consumption (including smokeless tobacco) on CAD risk \cite{Piano2010}. }
\item{\textit{Sedentary lifestyle:} Physical inactivity is a major contributor to risk for CAD and other chronic diseases \cite{Warren2010}, including obesity, hypertension, diabetes, and metabolic syndromes, which also increase risk for CAD \cite{Hamilton2007}. Among individuals with low physical activity, CAD risk is 2.7-fold higher than in individuals with higher physical activity  \cite{Hamilton2007}.}
\end{itemize}
}
\section{Genetics of Rare and Common Diseases}
\subsection{The Genome}
{There are 23 pairs of chromosomes in the humane genome: 22 pairs of autosomal chromosomes and one pair of sex chromosomes. The building blocks of DNA (deoxyribonucleic acid) are pairs of four bases (A, C, T, and G ) that form pairs A-T and C-G. Genes are stretches of bases in the DNA that are transcribed into messenger ribonucleic acid (mRNA) that give rise to proteins (i.e. central dogma of biology DNA{\textrightarrow}mRNA{\textrightarrow}proteins - which states that DNA determines how and which proteins are produced inside every cell, first through transcription then through translation into proteins). The number of protein-coding genes is estimated to be around 20,000. The protein-coding genes are crucial being the building blocks of all cells and tissue functions. They only form about 1.5\% of the DNA; the remaining 98.5\% of so called non-coding DNA is poorly understood but increasingly believed to have important gene regulatory roles.}
\subsection{Genetic Variation}
{Human chromosomes are paired, with one chromosome inherited from each parent. Any two individuals have 99.9\% similar sequence in their genome  \cite{Feuk2006}. The 0.1\% difference constitutes the genetic variation. Variable part of the genome covers an estimated 10-30 million bps that is found in at least 1\% of the population. Genetic variations arise from spontaneous mutations or recombination that remain conserved through evolution \cite{strachan2011human}. A location in the DNA with genetic variance that is associated with disease or other types phenotypes is called a genetic locus. Genetic variance in these loci is mainly from two types of DNA variants; copy number variations (CNVs) and single nucleotide polymorphisms (SNPs). CNVs are DNA sequences that appear at a variable number of copies. SNPs, the most common type of genetic variation, are single substitute, insertion, or deletion of base pairs (bp). On average there is one SNP in every 300 base pair. Each SNP has two alleles. Individuals with the two identical alleles are called homozygous, and those with two different alleles are called heterozygous.}\newline\indent
{Nearby SNPs tend to covary with each other. A set of SNPs that co-vary is called a haplotype block \cite{Manolio2008}. Two SNPs are said to be in linkage disequilibrium (LD) when their alleles are highly correlated in the population. A high LD corresponds to a statistical correlation, $r^2 \geq 0.8$ or $r^2 \geq 0.9$. A typical LD-block spans 60--200 kbp  \cite{Manolio2008,Hapmap2005,hapmap2007}.}
\subsection{Mendelian Diseases}
{Mendelian diseases are due solely to genetic causes (i.e., disease-causing variants) and follow a Mendelian inheritance pattern - one that is controlled by a single or few locus that is transmitted from parents to offspring. There are two types of Mendelian inheritance patterns: (1) autosomal dominant (or sex-linked dominant) inheritance, which requires only one copy of a defective allele or a set of alleles occuring in every generation (e.g., Huntington’s disease, or fragile X syndrome) \cite{Haines2006} The second one (2) is autosomal recessive (or sex-linked recessive) inheritance, which requires two copies of the same defective allele(s) (e.g., cystic fibrosis) or if on the X chromosome, one copy of the defective allele(s) for males or two copies of defective allele(s) in females (e.g., hemophilia A)  \cite{Haines2006}. Mendelian diseases can range from rare diseases caused by a set of variants in a specific locus to diseases caused by a single defective gene  \cite{Hoischen2010,Riordan1989,Kerem1989} to multigenic diseases caused by many genes  \cite{Haack2010}. An important aspect of Mendelian diseases is that the disease-causing variants have low frequency but high penetrance and a large effect size  \cite{Manolio2009}.}
\subsection{Common Complex Diseases}
{Common complex diseases (CCDs) are governed by inheritance patterns-heritable traits with a complex genetic architecture. The disease-causing variants found so far for CCDs, have, unlike Mendelian disorders, a higher frequency (minor allele frequency [MAF] $\geq$ 5\%), low penetrance, and a small effect size. Disease risk is in contrast to Mendelian diseases  thought to be caused by the combination of many genes also interacting with environmental factors. An important hypothesis for studies of common diseases is the “common disease, common variant” hypothesis, which assumed that common genetic variants (MAF $\geq$ 1--5\%) are the main contributors to genetic susceptibility of common diseases. However, recently by GWAS identified common variants explain about 10\% of disease heritability, and the majority of heritability in the population is unaccounted for  \cite{Manolio2009,Schork2009,Bodmer2008, Bjorkegren2015}. Some of the approximately 90\% missing heritability will be found by rare variants causing CCDs \cite{Manolio2009}. Other reasons might be so called epistatic effects, which roughly means that genetic variants linked to a given CCD act in conjunction to explain more of the genetic variance of the disease than the sum of their individual effects \cite{Wei2014}.}
\subsection{Design for Genetic Studies of Rare versus Common Complex Diseases}
{There are two main approaches for studying rare and common diseases:
\begin{enumerate}[label={\alph*)}]
\item {\textbf{\textit{Genetic linkage studies:}} In Genetic linkage studies identified genomic regions or gene(s) are linked based on patterns of inheritance  in families   \cite{Hirschhorn2005}. This approach has been dominant for rare diseases.}
\item {{\textbf{\textit{Genome-wide association studies:}} In Genome-wide association studies (GWAS) SNP alleles are tested for association with a disease phenotype at the population level   with a case--control setup\cite{Lewis2012}. This approach is used for genetic studies of CCDs. Two types of SNP-disease association can be inferred from GWAS: direct and indirect [Figure \ref{fig:IndAssoc}]. In a direct association, the SNP is the true causal variant for disease risk  \cite{Lewis2012,Cordell2005}. In an indirect association, the SNP is in LD with the true causal SNP  \cite{Lewis2012,Cordell2005}. SNPs can be in noncoding regions or protein-coding regions of the DNA. Most SNPs identified so far by GWAS are in in none-coding region and thus, believed to effect gene expression regulation rather than function of the gene.}
\begin{figure}
\captionsetup{width=0.95\textwidth}
\centering
\includegraphics[scale=0.7]{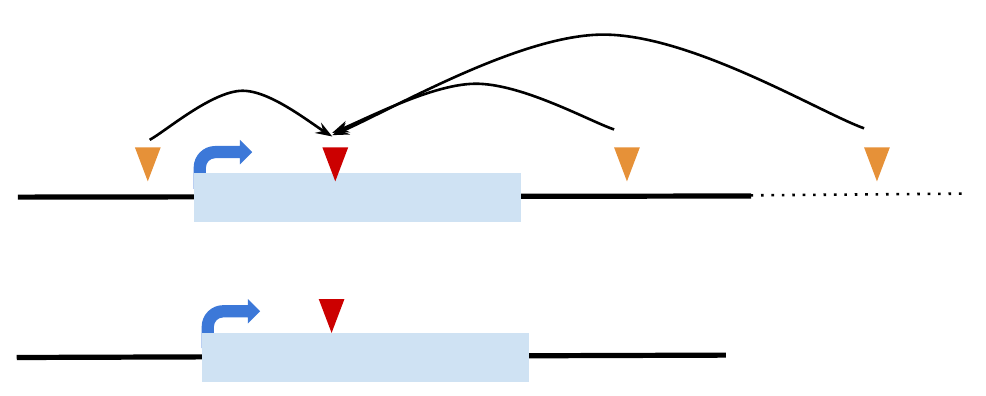}
\caption{\textit{Direct and indirect association.} If a genetic marker (orange) is in LD with a disease-susceptible allele (red), and there is evidence of association between the genetic marker and the disease phenotype, it is said that the disease-susceptible allele is also associated with the disease phenotype. This is known as ``guilt by association'' or indirect association.}
\label{fig:IndAssoc}
\end{figure}
\newline
{A main issues with GWAS is the multiple testing problem. Typically, 2 million SNPs are tested one-by-one resulting in that a p value of $2.5\times10^{-8}$ (after Bonferroni correction,) becomes the threshold to filter for genome-wide significant associations. A strategy to meet the multiple testing problem is to combine multiple GWAS datasets as a meta-analysis to increase power of the study.}
}
\end{enumerate}
}
\subsection{Genetics of Gene Expression Studies}
{
Systems genetics based on the use of genetics of gene expression studies (GGES) is an increasingly appreciated solution to complement GWAS for better understanding of CCD etiology. GGES introduces one or several intermediate phenotypes between DNA and the phenotypes shortening the biological distance allowing for smaller studies in the range of hundreds to a thousand individuals. We focused on gene and RNA expression profiles, but future studies will also consider other intermediate phenotypes, such as protein and metabolite expression  \cite{Schadt2012}. RNA is a useful first step in mapping expression quantitative trait loci (eQTLs) across the genome and to decipher gene networks  \cite{Schadt2009}. Detecting eQTLs is a powerful technique to assign gene regulatory functions of DNA variants. The fact that gene or RNA expression is an immediate consequence of DNA genotypes make these associations statistically stronger than for those sought between DNA and phenotypes in GWAS - allowing to detect eQTLs in smaller studies (i.e. smaller number of study subjects) than what is needed in GWAS.}\newline \indent
{
This section will start by describing quality control (QC) procedures for GGES and by explaining how to map eQTLs. It will conclude by covering kruX and MatrixeQTL, the two eQTL mapping methods we used in this thesis.
}
\subsection{Quality Control for Genotype Data}
{
Several measures must be taken into account to ensure that the data are of sufficient quality to prevent false  variation frequently related to technical noise from sample handling rather than true biological or disease variation – this is done during the quality control (QC) procedure. Here are some QC steps we used in this thesis:

\begin{itemize}
\item{\textit{\textbf{Individual gender check:}} It is important to filter and QC for gender mismatches, as it can be a covariate in a study. This QC step can be done at the genotype level, usually method by checking for SNPs on X chromosomes. Since X chromosomes are typically genotyped for SNPs, all Y chromosome SNPs will be missing. Thus males should have a homozygosity rate of 1 and females should typically have a much lower homozygosity rate  \cite{Anderson2010}.}

\item{\textit{\textbf{Individual genotype call rate:}} A very high threshold for genotype call rate can affect the data in two ways: loss of power from loss of individuals being filtered out and “informative missingness”. The former can happen if DNA sample quality is not high enough, which leads to most SNPs being improperly called. The latter is more complex and is a dilemma for genotype-calling algorithms. These algorithms attempt, on the basis of signal intensity, to assign a genotype with very high certainty or to “guess” a genotype with a low certainty. The former may lead to ``informative missingness", where a genotype might be correlated to its status of being missed (e.g. a rare genotype) \cite{Anderson2010}. A recommended 3--7\% call rate threshold has been previsouly used  \cite{Burton2007}.}
\item{{\textbf{\textit{Individual relatedness:}} Generally speaking, related (or duplicate) individuals can introduce false positive or false negative results (by affecting allele frequencies). Two measures are used to address this problem: identity by state (IBS) for duplicates and identity by descent (IBD) for relatedness between individuals. IBS shows how similar to each other two or sets of alleles or portions of DNA in two individuals are; IBD shows how related these segments are in terms of sharing common ancestry  \cite{Anderson2010,Gillespie2010}.}{First, the short and long range of LD should be addressed to remove highly correlated SNPs. The reason is that individual relatedness analysis achieves the best results by assuming that SNPs are not in LD with each other. Sometimes, the “extended LD ” regions are removed entirely  \cite{Anderson2010}. The rest of the SNPs within short-range LD regions should be LD-pruned. \cite{Anderson2010}.}
{Second, the IBS and IBD scores should be analyzed  \cite{Purcell2007}. If two individuals have very high IBS score, close to 1, they are most probably duplicates. And from genome-wide IBS data, we can derive IBD. Duplicates or monozygotic twins have an IBD score of 1, first-degree relatives have an IBD score of 0.5, second-degree relatives an IBD score of 0.25, and third-degree relatives an IBD score of 0.125. IBD scores can deviate from these values due to population structures and genotyping errors. For example, two siblings that share same mother but two different fathers (who are brothers) will have an IBD score of 0.3125 (0.25 + 0.125/2). However, an IBD score of 0.1875 is used to remove related individuals (mean score between second- and third-degree individuals) \cite{Anderson2010}.}}

\item{\textbf{\textit{SNP genotype call rate:}} Removing low-quality SNPs and those with a low genotype call rate will help reduce both false positive and false negatives. Different measures are taken here. Generally, a 95\% call rate is considered the threshold for filtering out SNPs  \cite{Anderson2010}.}

\item{\textbf{\textit{SNP MAF:}} MAF is dependant on sample size \cite{Neale2008, Anderson2010}. SNPs with very low MAF ($\leq$1-–5\%) tend to end up as false positives. For a very large dataset, a threshold of 1-–5\% is enough to ensure sufficient power to detect rare variants. In a smaller dataset, however, it is necessary to observe at least 20–30 of a given genotype  \cite{Neale2008}. A rule of thumb is to put the MAF threshold at 10/n, where n is the sample size. This way SNPs that pass this threshold, might have the chance to pass the genome-wide significance level in case control studies. For example, in a GWAS setup where we have cases and controls, if a SNP is present only in 25 heterozygous, and all of them in cases, the \textit{P} value will be approximately $0.5^{25}\approx 2.98\times 10^{-8}$.}

\item{\textbf{\textit{Hardy-Weinberg equilibrium (HWE):}} Deviation from HWE is associated with false positives \cite{Anderson2010,Neale2008}. Such deviation can happen when genotype call algorithms do not assign genotypes properly. Generally, the p value threshold for HWE is set at relaxed threshold of $10^{-3}$ or a stringent threshold of $10^{-6}$  \cite{Burton2007,Neale2008}.}

\end{itemize}
}
\section{eQTL Mapping}
\begin{figure}
\captionsetup{width=0.95\textwidth}
\centering
\includegraphics[scale=0.5]{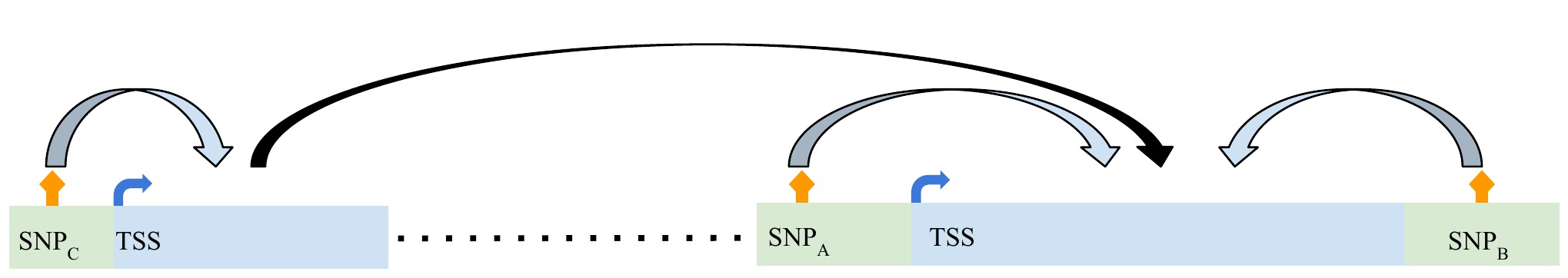}
\caption{\textit{Cis- and trans-eQTLs.} Cis-eQTLs are eQTL within a certain distance of transcript start (TSS) and transcription end sites (TES) (e.g. $_{A}$ and SNP$_{B}$ within 1 Mbp ). Trans-eQTLs are more than 1 Mbp from TSS and TES, and they can also reside on different chromosomes than their regulating genes (e.g. SNP$_{C}$).}
\label{fig:eQTLMap}
\end{figure}

{GGES use the power of intermediate phenotypes such as mRNA abundance as a sensor between genome and a clinical phenotype. Genotypes are integrated with mRNA abundance (i.e., gene expression) to link the genotypes with gene expression. In one of the first attempts  \cite{Brem2002}, the global gene expression of 6215 genes was compared against 3312 SNPs in 40 yeast segregants. Essentially, this was done by using two matrices: a 6215 $\times$ 40 matrix for gene expression and a 3312 $\times$ 40 matrix for SNPs. In this study eQTLs were defined as cis- and trans-acting eQTLs; Cis-acting eQTLs (cis-eQTLs) as SNPs within 10 kbp of gene boundaries and trans-acting eQTLs (trans-eQTL) as those outside this range [Figure \ref{fig:eQTLMap}].}
\newline \indent
{With increasing number of SNPs, individuals, and other quantities, there is increasing need to adopt methods that provide faster and accurate results. In this thesis, two different methods were used for eQTL mapping: kruX, and Matrix-eQTL. The former is described in paper I, and the latter was published by Shabalin in 2012  \cite{Shabalin2012}.}
\subsection{Permutations and Resampling}
{
Permutation tests are nonparametric randomization procedures for determining statistical significance or estimating empirical false-discovery rates (FDR). These procedures are a robust approach in which data labels are shuffled multiple times. In each shuffled set (i.e., a permuted data set), test statistics are computed. The test statistics of permuted data sets generate a distribution under the null hypothesis. By aggregating these test statistics and comparing with the observed one, one can estimate the probability of obtaining a similar or an extreme effect compared to the null hypothesis. Permutation tests are specifically useful when the null hypothesis is “no association”. For eQTL mapping and estimating FDR, sample labels either in genotype or gene expression data matrices are permuted. This approach was used in paper I to estimate empirical FDR. For inherited risk enrichment analysis, a resampling procedure was adopted to mimic the experimental data set, as explained below and it was used in papers II and III, accordingly.}
\section{Inherited Risk Enrichment Analysis}
{Gene and SNP set inherited risk enrichment analyses have been gaining attention during past years  \cite{Weng2011, Subramanian2005}. These methods attempt to analyze the association of group of functionally related genes with a disease of interest as opposed to traditional analysis of GWAS data focusing on SNPs one-by-one. In this fashion, GGES may compensate for potential drawbacks of GWAS that may underlie the fact that genome-wide significant variants so far explain only a small portion of CCD heritability  \cite{Manolio2009}. Some of the weaker associations (i.e. $P\geq10^{-8}$) filtered out by GWAS because of the vast multiple testing correction may in fact be true variants affecting disease. Gene and SNP set risk enrichment analysis methods instead focus on testing prioritizing and grouping signals gathered from multiple SNPs either defined by genes  \cite{Subramanian2005}, pathways  \cite{Zhong2010}, or by selecting SNPs based on their effect on gene regulating in the form of eQTLs  \cite{Schadt2008}.}
\subsection{From SNPs and Genes to eQTLs}
{A meaningful link between inherited genetic risk between genes or SNPs and disease should be established. This is crucial to be able to explain the proportion of inherited risk that is not explained by GWAS. Thus, in using GWAS results, an important step to analyze SNPs or genes for inherited risk for a disease is to make a link between the experimental set of genes to SNPs in GWAS. If the starting set is a list of SNPs, this can be achieved using GGES and eQTLs. Basically, eQTLs for each SNPs will be selected either through a direct hit or using LD. $r^{2}$ which is the correlation between two SNPs and reflects the LD strength can used. The cutoff for $r^{2}$ can be the same as the cutoff used to identify so-called tag SNPs in HapMap  \cite{hapmap2007,Hapmap2005}. This step, to select and expand SNPs using eQTL datasets, is referred to as ``LD expansion". The starting set can also be a list of genes. This list can come from different sources, including computational approaches such as clustering approaches, or gene interaction network analysis, or it can come from annotated sources such as The Kyoto Encyclopedia of Genes and Genomes  \cite{Kanehisa2000}, Gene Ontology (GO)  \cite{TheGeneOntologyConsortium2000,TheGeneOntologyConsortium2015}, or REACTOME  \cite{Croft2014}.}
\newline
\indent
{SNPs can be assigned to genes in various ways. It can be done through the proximity of an SNP to a gene or through a functional or statistical association with the gene. For distance, different measures have been used, including 5 kb  \cite{Chen2010}, 20 kb  \cite{Jia2010}, 50 kb  \cite{VanDerLaan2015}, 200 kb  \cite{Perry2009}, and 500 kb  \cite{Wang2007}. Although SNPs that influence gene expression have been found as close as 20 kb from a gene  \cite{Veyrieras2008}, these types of selection for SNPs might also pick up those that are in no way associated with the gene. To overcome this problem, eQTLs from GGES studies is useful. eQTLs will provide a true link between a SNP- gene pair. A practical issue with eQTLs is that not all of the relevant SNPs will be selected. One way to solve this is to use the LD-expansion technique described above. This way, those SNPs that were not picked up by eQTL studies will be covered as well.
}
\subsection{From eQTLs to GWAS}
{The next step would be to identify how the LD-expanded set overlaps with the millions of SNPs identified through GWAS. This can be done by accessing GWAS summary results and capturing the disease association \textit{P} value for SNPs in the LD-expanded set. Since some genes might be close to each other, the identified SNPs (through eQTLs) might be redundant. This can be a potential source of bias due to LD between SNPs \cite{Wang2007, Cantor2010}. The effect of LD bias should be mitigated as much as possible to prevent inflation or deflation of signals. This can be done by LD pruning techniques to reduce the number of redundant SNPs. This set is called LD-pruned set of SNPs.}
\subsection{Statistical Significance Assessment}
\begin{figure}
\captionsetup{width=0.95\textwidth}
\centering
\includegraphics[scale=0.6]{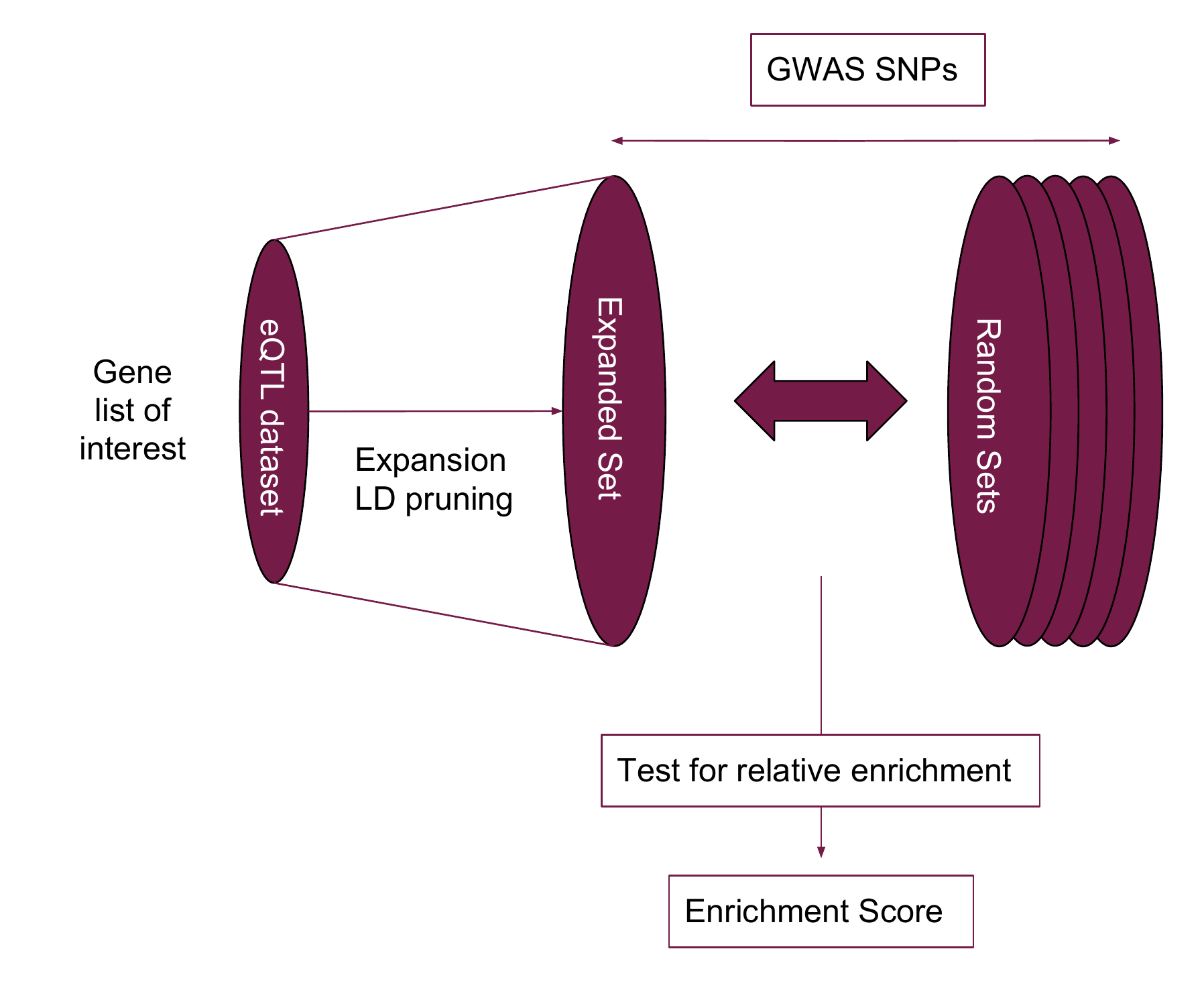}
\caption{\textit{Schematic of inherited risk enrichment analysis.} This schematic illustrates an overview of inherited risk enrichment analysis. Starting from a set of genes to finally assign an enrichment score to the gene set.}
\label{fig:EnrichScheme}
\end{figure}
{To assess statistical significance, a resampling technique can be used. If the starting set is enriched with disease causing SNPs or genes, the resulting set (after pruning) should also be enriched in disease risk. To test this, an enrichment score can be defined as the ratio of disease-causing SNPs to non-disease-causing SNPs compared to the background data, which is millions of SNPs in the GWAS summary results. The SNPs in a GWAS panel, however, do not reflect the smaller set of SNPs in the LD-pruned set of SNPs. Millions of SNPs in GWAS panel represent the whole genome, but the SNPs in the LD-pruned set might be distributed only in parts of genome or even aggregate on a single chromosome. So a null distribution must be constructed to mimic the LD-pruned set as much as possible. A couple of factors should be considered. First, we must consider the chromosomal distribution of SNPs. A background data set should have the same number of SNPs in each chromosome as the LD-pruned set. Second, we must consider MAF distribution. Again, it is important to construct background data based on the MAF distribution of LD-pruned set. To do so, SNPs in the LD-pruned set are binned into 5 groups. These 5 groups are starting from 0 with steps of 0.1. Third, we must select those SNPs based on the criteria for the LD-expanded set. That is, if SNPs were selected based on cis-eQTLs that are within 1 Mbp of the gene window, the background data should only include those SNPs from GWAS summary result that are within 1 Mbp of a gene. Last but not least, the number of SNPs in the background data should match number of SNPs in the LD-pruned set. This procedure should be repeated a couple of thousand times to approximate a null distribution. All the procedures above are proposed as an algorithmic framework under the term inherited risk enrichment analysis as described in Figure \ref{fig:EnrichScheme} and Algorithms 1 and 2.}

\begin{algorithm}[H]
\caption{Stage one: Assign $F_{real}$ to the list of input genes}
\begin{algorithmic}
    \STATE {\textbf{Input:} A list of genes}
    \STATE {\textbf{Output:} $F_{real}$}
    \bindent
    \begin{enumerate}
		\STATE {Map eQTLs for each gene in the gene set from GGES dataset}
		\STATE {Expand eQTLs using LD}
		\STATE {Prune the expanded eQTL list}
		\STATE {Record disease association \textit{P} values for each SNP in the pruned set from GWAS dataset}
		\STATE {$N_{real} \gets$ SNPs with \textit{P} value $\leq 0.05$}
		\STATE {$N_{tot} \gets$ Total number of SNPs with a disease association \textit{P} value}
		\STATE {$F_{real} \gets \frac{N_{real}}{N_{tot}} $}
    \end{enumerate}
    \eindent
\end{algorithmic}
\end{algorithm}

\begin{algorithm}[H]
\caption{Stage two: Resampling procedure to construct a null distribution for $F_{scr}$}
\begin{algorithmic}
    \STATE {\textbf{Input:} List of LD pruned SNPs from stage one}
    \STATE {\textbf{Output:} $F_{rand}$}
    \bindent
    \begin{enumerate}
	    \STATE {If using \textit{cis-}eQTLs, select only SNPs within 1Mb of TSS of genes from GWAS summary results}
    		\STATE {Permute SNP index IDs in GWAS summary results}
    		\STATE {For n=1,...,10000 select $N_{tot}$ SNPs by:}
    		\begin{enumerate}
			\STATE {Matching chromosomal distribution of LD pruned eQTLs}
			\STATE {Matching MAF distribution of SNPs (five MAF bins $[0,0.1),...,[0.4,0.5]$)}
			\STATE {$N_{rand}^{n} \gets$ SNPs with \textit{P} value $\leq 0.05$}
			\STATE {$F_{rand}^n \gets \frac{N_{real}^{n}}{N_{tot}} $}
    		\end{enumerate}
    		\STATE {$F_{scr} \gets \frac{F_{real}}{\langle{F_{rand}}\rangle}$}
    		\STATE {Calculate \textit{P} value $P(Z\leq z)$: \newline
    		$z=\frac{N_{real}-\bar{N}_{rand}}{\sqrt{\frac{1}{1-n}\sum_{i=1}^{n}{\vert{N_{rand}^{i}-\bar{N}_{rand}}^{2}}}\vert}\quad,\ n=10000$}
    \end{enumerate}
    \eindent
\end{algorithmic}
\end{algorithm}

\begin{center}
\decoone
\end{center}


\chapter{Aims}
{
The overall aim of this thesis is to develop a reliable method for eQTL mapping from genetics of gene expression studies using the STockholm Atherosclerosis Gene Expression (STAGE) study of coronary artery disease (CAD) patients allowing to a build a computational pipeline for inherited risk enrichment analysis of eQTLs by re-examining datasets of genome-wide association studies (GWAS). New and existing methods were used to develop these methods to assess the heritability (and thereby causality) of eQTLs and associated genes active in molecular processes (in up to regulatory gene networks) identified in specific and multiple tissues important for primarily CAD.}
\section{Specific Aims}
{The specific aims of the individual papers are:
\begin{description}
[align=left]
\item{\textit{Paper I:} To develop a fast nonparametric eQTL mapping method.
\item{\textit{Paper II:} To develop a method for inherited risk enrichment analysis based on eQTLs and GWAS datasets and to use this method to investigate the inherited risk enrichments of eQTL sets identified in seven CAD-relevant tissues of the STAGE study.}
\item{\textit{Paper III:} To apply the method of inherited risk enrichment analysis to multi-tissue regulatory gene networks identified in STAGE.}
\item{\textit{Paper IV:} To use the STAGE and Athero Express studies to map eQTLs of three atherosclerosis candidate genes (ALOX5, ALOX5AP, and LTA4H) and their associations with phenotypes in Athero Express.}
}
\end{description}

\begin{center}
{~}
\newline
\decoone
\end{center}


\chapter{Present Investigations}
\section{Paper I: kruX an eQTL Mapping Tool}
{mRNA abundance is one of the most proximal sensors of DNA variation. It reflects the interactions between the genotype and microenvironment within a tissue  \cite{Schadt2012,Schadt2009}. This interaction can be captured at any stages of the disease. eQTL mapping tools can be used to analyze combinations of hundred thousands of SNPs and tens of thousands of genes (or any qualitative trait) all at once. In this paper, we developed a bioinformatic tool for fast eQTL mapping. We named it kruX, which is based on the Kruskal-Wallis test statistics. In summary, kruX excels in detecting higher proportion of non-linear associations than other methods. This tool was developed for Python, MATLAB, and R, and it is available online.}
\begin{figure}
\captionsetup{width=0.95\textwidth}
\centering
\includegraphics[scale=0.5]{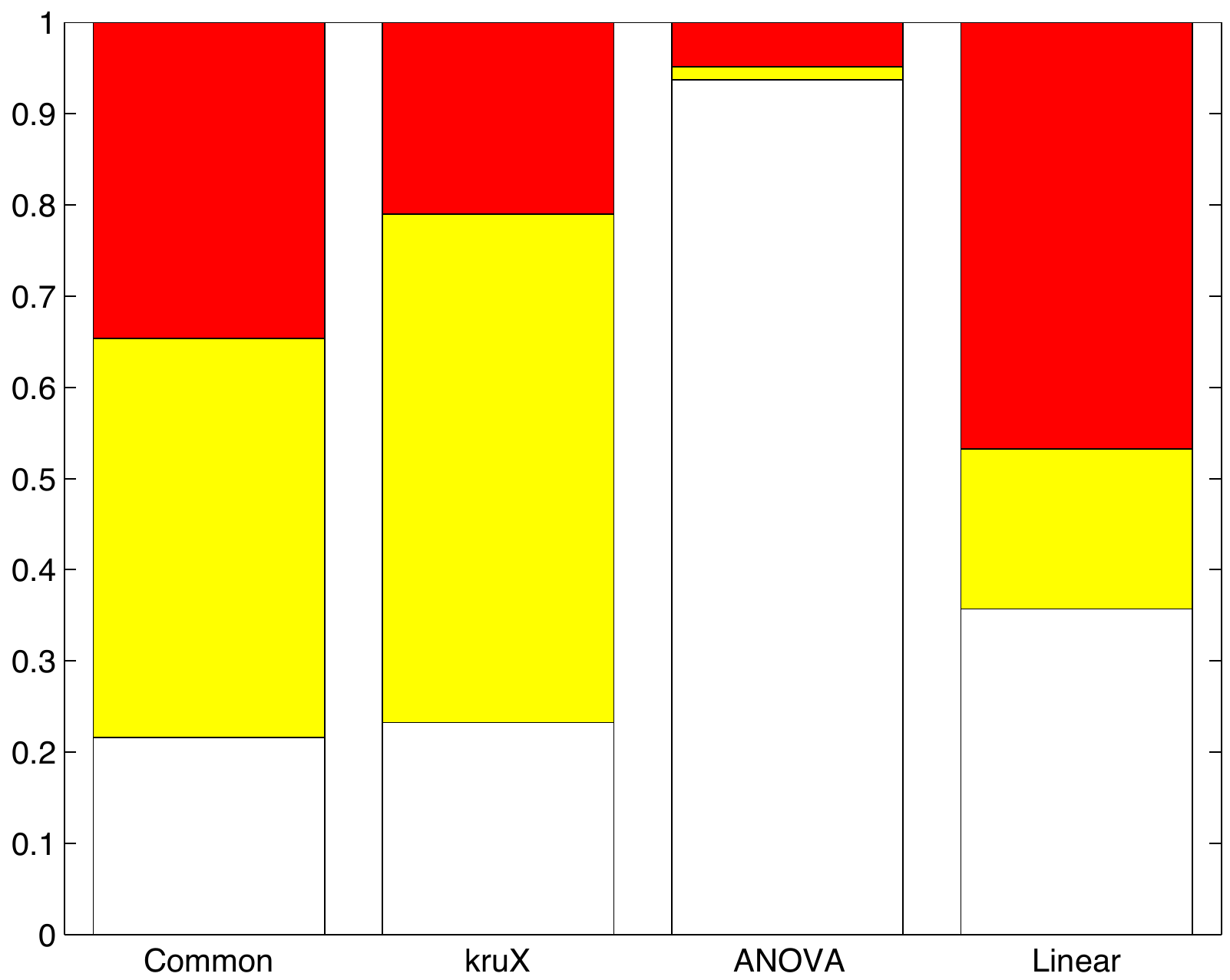}
\caption{\textit{Relative proportion of identified eQTLs and their types.} Identified eQTLs are divided into three groups: common eQTLs (white), skewed genotype group (yellow), and nonlinear eQTLs (red). Each bar represents one group of eQTLs. Common refers to the eQTLs found by all three models. kruX has the highest relative proportion of identified nonlinear associations, while linear and ANOVA models are second and third, respectively}
\label{fig:kruXResult}
\end{figure}

\subsection{Matrix Multiplications for Fast Calculations}
{kruX has two input matrices: genotype (G) and gene expression (D). G is organized as M genotype marker by K individuals; D is organized as N transcripts by K individuals. In kruX, genotype markers have the values 0, 1, or 2 (i.e., AA, Aa, and aa). This method assumes that the data has passed necessary QC filters. kruX will convert matrix G into a sparse logical matrix, and rank gene expression matrix D into a rank matrix of R. Then the Kruskal-Wallis test statistics matrix S will be calculated through matrix multiplications. For \textit{P} value calculations, a nominal \textit{P} value threshold of $P_{c}$ is chosen by the user. Corresponding test statistics threshold for $P_{c}$ is then calculated for 1 degree of freedom. kruX then will remove those that exceed test statistics threshold from matrix S. \textit{P} values for the rest of the values are then calculated using $\chi$\textsuperscript{2} distribution. Empirical FDR values are estimated by permuting columns on the expression data ranks. The FDR value for is then defined as ratio of average number of associations in the permuted data to the number of associations in the original data. kruX comes with an example data and script from 2000 randomly selected genes and markers from 100 randomly selected yeast segregates  \cite{Brem2005}.}
\subsection{kruX is Fast and Accurate}
{kruX was validated by testing each and every association one by one using built-in Kruskal-Wallis functions. We then analyzed kruX for 19,610 gene expression profiles and 530,222 SNPs from the STAGE dataset  \cite{Hagg2009}. kruX relies on matrix multiplication to calculate test statistics and was at least 11,000 times faster than running a Kruskal-Wallis test for each SNP-gene association test. We compared the results from kruX to the results from the popular Matrix-eQTL method  \cite{Shabalin2012}. Matrix-eQTL has two parametric ANOVA and linear model options. Since the Kruskall-Wallis test is more conservative than ANOVA and linear models, the nominal \textit{P} values will be higher. Thus it is not possible to compare \textit{P} values directly. To be able to compare them, we performed empirical FDR correction for multiple testing and filtered out identified eQTLs at various FDR cutoffs. Most of the identified eQTLs didn’t pass the FDR filter when we used the ANOVA model in Matrix-eQTL. The sheer numbers of eQTLs identified with kruX and with the linear model in Matrix-eQTL, however, were more or less comparable. We found that this is due to pairs of rare homozygous minor alleles and gene expression outliers. However, the results were more or less comparable when considering cis-eQTLs only. ANOVA in Matrix-eQTL also proved to be sensitive to gene expression outliers and SNPs with rare homozygous alleles. The linear model showed the highest number of associations after FDR correction, owing to the additive linear associations. The eQTLs identified with all three approach are summarized and compared in Figure \ref{fig:kruXResult}.}
\section{Paper II A Multitissue Repository of eQTLs for CAD}
{GWAS have identified at least 150 loci for CAD and MI  \cite{Welter2014}. Of these 150, 46 were found through GWAS meta-analysis  \cite{Peden2011,Schunkert2011,Deloukas2013}. These loci are believed to contribute to at least 10\% of inherited risk  \cite{Deloukas2013}. SNPs that do not reach a genome-wide significance level after \textit{P} value correction ($\geq10^{-8}$) and rare SNPs might contribute to the missing ~90\% of heritability  \cite{Schadt2012,Orozco2012}. Since CAD is a complex systemic disease, disease-causing variants and SNPs might exert their risk for CAD across multiple tissues  \cite{Tegner2007,Tegner2007b}. In this paper, genome-wide genotype and global gene expression data from the multi-tissue STAGE study were used to identify eQTLs in and across seven CAD-relevant tissues  \cite{Hagg2009}. In summary, we identified 16 multi-tissue eQTLs associated with 19 master regulatory genes. Three of the 19 master regulatory genes segregated the patients according to the extent of CAD. Furthermore, siRNA targeting of these genes in the THP-1 monocytic cell line also incubated with acetylated LDL affected cholesterol-ester accumulation in this in vitro model of foam cell formation \cite{Skogsberg2008}.}
\begin{figure}
\captionsetup{width=0.95\textwidth}
\centering
\includegraphics[scale=0.2]{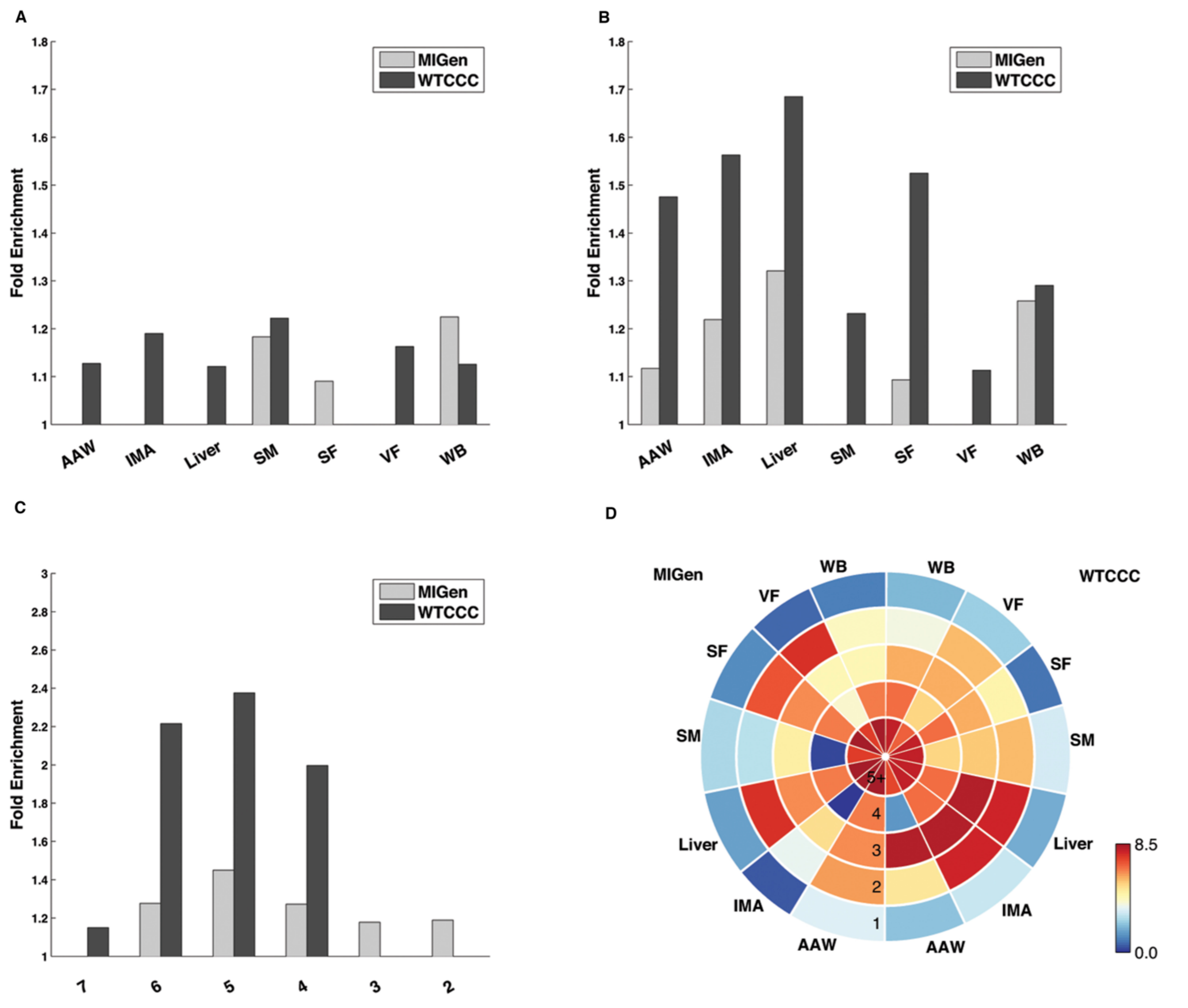}
\caption{\textit{Inherited risk enrichment analysis results for cis-eQTLs in STAGE.} (A) CAD/MI risk enrichment for tissue-specific eQTLs. (B) CAD/MI risk enrichment for tissue shared eQTLs. (C) CAD/MI risk enrichment for cis-eQTLs in found in 2 to 7 tissues. (D) Pie chart showing risk enrichment for eQTLs in tissue combinations. Each pie represents one tissue. Each layer for each pie shows the enrichment of tissue shared eQTLs and a combination of 1 to 5 tissues. For each section in each pie, the highest risk enrichment score for the possible combinations of the said tissue with other tissues is indicated.}
\label{fig:EnrichResult}
\end{figure}
\subsection{Integrating GGES and GWAS}
{Samples of seven CAD-relevant tissues - atherosclerotic arterial wall (AAW), internal mammary artery (IMA), subcutaneous fat (SF), visceral fat (VF), skeletal muscle (SM), liver, and whole blood (WB) - were obtained from STAGE patients for RNA extraction. Whole-blood samples were also obtained for RNA and DNA extraction. DNA from 109 patients was genotyped with GenomeWideQTL\_6 arrays (Affymetrix).
Custom-made HuRSTA-2a520709 arrays (Affymetrix) were used for gene expression profiling of the RNA samples. kruX was used to survey 19,610 identified mRNA expression traits for association with each of the genotyped SNPs at FDR 15\%. cis-eQTLs were defined within 1Mb of transcription start or end site, and the rest of eQTLs as trans-acting eQTLs. The cis-eQTL with lower \textit{P} value for each gene in each tissue was used for the inherited risk enrichment analysis. The cis-eQTLs were LD-expanded with SNP at $r^2$ threshold of 0.8 according to the European population panel in 1000 Genomes project data \cite{1000genome2010}. Next, the summary statistics of GWAS datasets from two independent studies and one meta-analysis were  used to assign the expanded cis-eQTL group with \textit{P} values indicating associations with CAD. For this we used two independent GWAS: the Wellcome Trust Case Control Consortium (WTCCC)  \cite{Burton2007} and the Myocardial Infarction Genetics Consortium (MIGen)  \cite{Kathiresan2009} studies. For replication and validation of results, we used Coronary Artery Disease Genome-wide Replication and Meta-analysis (CARDIoGRAM)  \cite{Schunkert2011,Deloukas2013}}
\newline
\indent
{Next, sets of genes co-expressed with genes regulated by cis-eQTLs in the same tissue were identified at a Pearson correlation coefficient threshold $\geq0.85$. These gene sets were then characterized using Gene Ontology (Fisher's exact test). Benjamini and Hochberg FDR corrections were used to correct for multiple testing. 
}
\begin{table}[]
\centering
\captionsetup{width=0.95\textwidth}
\caption{Cis-eQTLs identified in seven different tissues and their corresponding number of genes and \textit{P} value cutoff}
\label{my-label}
\begin{tabular}{@{}lllll@{}}
\toprule
Tissue & \MyHead{1.5cm}{No. of eQTLs} & \MyHead{2.5cm}{No. of Genes} & \MyHead{2.5cm}{No. of Gene- eQTL Pairs} & \MyHead{2.5cm}{\textit{P} value cutoff} \\
\midrule
\MyHead{3cm}{AAW} & 3579 & 799 & 3716 & $2.5\times10^{-4}$ \\
\MyHead{3cm}{IMA} & 4979 & 1122 & 5304 & $2.9\times10^{-4}$ \\
\MyHead{3cm}{SF} & 3867 & 876 & 4024 & $3.0\times10^{-4}$ \\
\MyHead{3cm}{VF} & 4012 & 923 & 4324 & $2.4\times10^{-4}$ \\
\MyHead{3cm}{SM} & 5046 & 1095 & 5315 & $3.0\times10^{-4}$ \\
\MyHead{3cm}{Liver} & 10180 & 2158 & 10927 & $4.0\times10^{-4}$ \\
\MyHead{3cm}{WB} & 15012 & 2984 & 16952 & $6.7\times10^{-4}$ \\
\MyHead{3cm}{Total cis-eQTL} & 29530 & 6450 & 34611 & N/A \\
\MyHead{3cm}{Total shared cis-eQTL} & 7429 & 1839 & 6986 & N/A \\
\MyHead{3cm}{Total tissue specific cis-eQTL} & 22101 & 4611 & 27625 & N/A \\ \bottomrule
\end{tabular}
\end{table}

\subsection{Patterns of Carrying Disease Risk for CAD in Groups of eQTLs}
{We identified 29,530 cis-eQTLs for 6,450 unique genes by analyzing 19,610 genes across 7 tissues (Table 4). 25\% of all the identified cis-eQTLs (7,490) were found in at least two other tissues. When investigating individual tissues, at least 60\% of cis-eQTLs in each tissue were also found in another tissue. These eQTLs were referred to as tissue-shared eQTLs. The total number of identified trans-eQTLs was 1,494, and 2.9\% of them were found in at least two tissues. Similar proportions for cis- and trans-eQTLs were previously reported  \cite{Greenawalt2011,Fairfax2012,Fu2012,Grundberg2012}. In general, tissue-shared eQTLs were found to be closer to transcription start and end sites than tissue-specific eQTLs.}
\newline \indent
{The entire set of cis-eQTLs was enriched in CAD risk by 1.15-fold in the MIGen GWAS data set  \cite{Kathiresan2009} and by 1.2-fold in the WTCCC data set  \cite{Burton2007}. In general, tissue-shared eQTLs in each tissue were more enriched in CAD risk than tissue-specific eQTLs [Figure \ref{fig:EnrichResult}A and \ref{fig:EnrichResult}B]. We found that eQTLs shared among 4, 5, or 6 tissues were more enriched than those shared among 2, 3, or 7 tissues [Figure \ref{fig:EnrichResult}C]. Investigating this further, tissue-shared eQTLs in specific combination of tissues showed a similar trend. In other words, we ran inherited risk enrichment analysis for all of the 120 possible combinations of tissues. The inherited risk enrichment of eQTLs peaked at combination of 5+ tissues in both GWAS datasets [Figure \ref{fig:EnrichResult}D]. Taking the eQTLs identified across at least 4 tissues, we found 42 multi-tissue eQTLs in two sets of 5 (AAW, liver, SF, VF, and WB) and 6 (IMA, liver, SM, SF, VF, and WB) tissue combinations to be highly enriched. The enrichment for these two sets were replicated in CARDIoGRAM, a GWAS meta-analysis for CAD  \cite{Schunkert2011,Deloukas2013}.}
\newline \indent
{We analyzed the above two sets by assessing the downstream effects of gene expression governed by identified 42 eQTLs. We found 29 gene sets for 16 of the 42 eQTLs. These 16 eQTLs were referred to as master regulator eQTLs. Nineteen unique cis-regulated genes (of 29 gene-sets) were also identified. These genes are referred to as master regulator genes. The identified 16 master regulatory eQTL set was enriched even further than the starting set of 42 eQTLs according to the CARDIoGRAM GWAS dataset. We assessed the functional characteristics of the 29 gene sets using Gene Ontology  \cite{TheGeneOntologyConsortium2015}. 19 gene sets had a significant \textit{P} value for CAD-relevant categories. These gene sets were also associated with CAD phenotypes in STAGE. Four of the master regulatory genes were associated with both functional and biological processes and with the degree of CAD in STAGE. The disease association of these genes was further validated in a THP-1 foam cell model by siRNA targeting. siRNA inhibition of three of the four (G3BP1, FLYWCH1, and PSORS1C3) master regulators markedly reduced cholesteryl-ester accumulation in a THP-1 foam cell model.}

\section{Paper III: A Cross-tissue Analysis of Regulatory Gene Networks}
{Molecular networks are crucial for understanding how genetic and environmental factors interact to affect CCD development and progress. To achieve a clear understanding of molecular and regulatory gene landscape, an integrative systems genetic approach was proposed  \cite{Bjorkegren2015,Civelek2014,Schadt2012}. This is done by identifying co-expression modules, inferring regulatory gene networks (RGNs) and by identifying key drivers in these RGNs. In this paper, a multi-tissue cross-species systems genetic approach was taken to identify and study RGNs and their role in CAD. First, co-expression modules were identified by weighted gene co-expression network analysis (WGCNA) across seven tissues. Then, to investigate the CAD causality for these modules, we used the CARDIoGRAM datasets for the inherited risk enrichment analysis. RGNs were inferred by Bayesian framework and key driver analysis. The cross-species conservation of the identified RGNs was investigated using the Hybrid Mouse Diversity Panel (HDMP)  \cite{Bennett2010}. Finally, key drivers in cross-species validated RGNs were further targeted with siRNA in the THP-1 foam cell model.}

\subsection{Identifying Cross-tissue Co-expression Modules Associated with CAD}

{Co-expression modules were identified across tissues using the WGCNA method  \cite{Langfelder2008}. 
\begin{figure}[htbp]
\captionsetup{width=0.95\textwidth}
\centering
\includegraphics[scale=0.7]{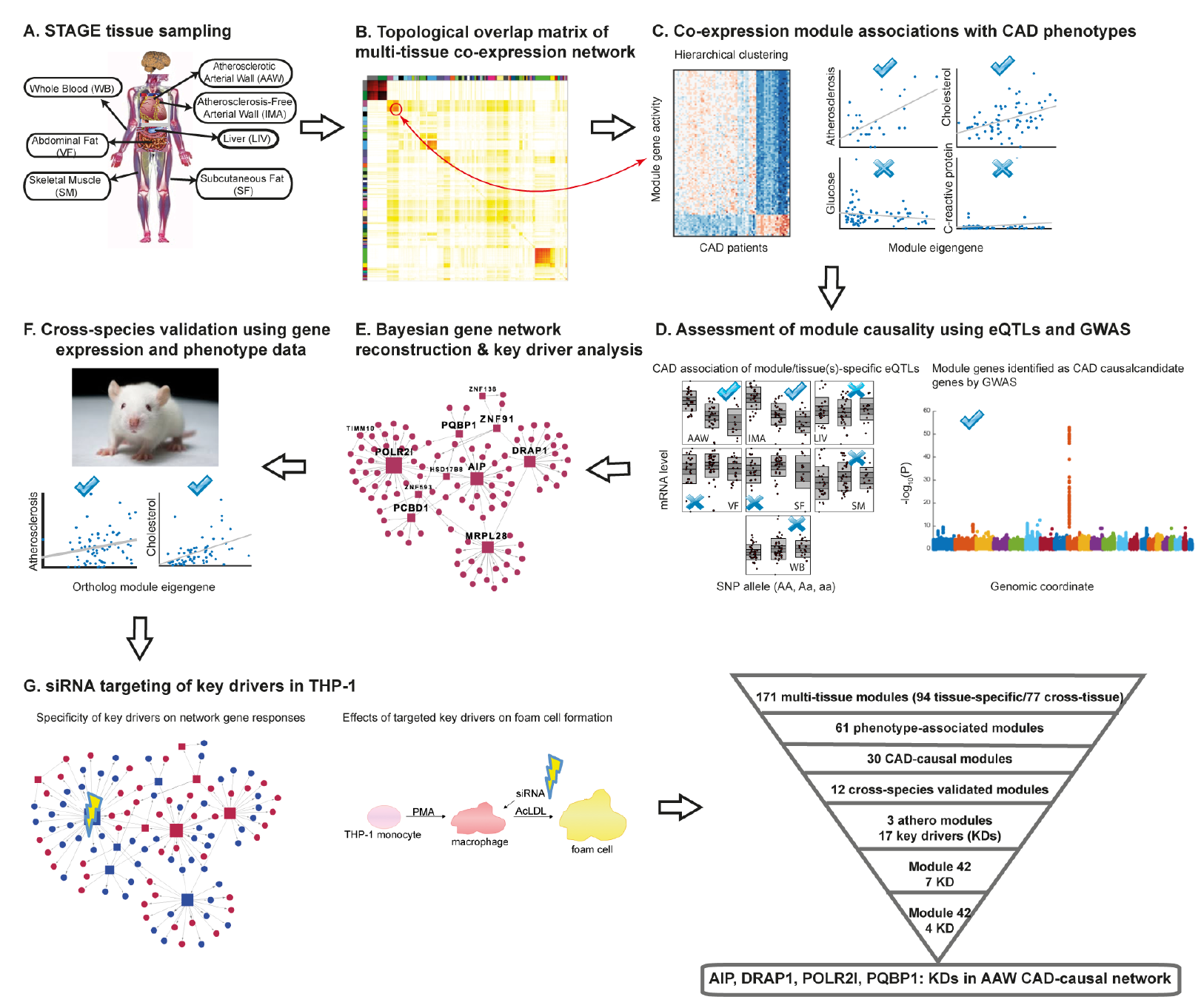}
\caption{\textit{Schematics of analytical steps starting with the STAGE study.} (A) Sampling tissues in STAGE study for mRNA profiling. (B) Multi-tissue WGCNA method for constructing tissue-specific and cross-tissue modules. (C) Detecting co-expression module associations with 4 main CAD phenotypes assessed in the STAGE study. (D) Analyzing causality of modules using STAGE eQTLs and GWAS using inherited risk enrichment and GSEA-like analyese. (E) Reconstruction of RGNs by using Bayesian framework and key driver analysis. (F) Using the HMDP dataset for cross-species validation of RGNs. (G) siRNA targeting of key drivers in the foam cell in vitro model.}
\label{fig:crosstissScheme}
\end{figure}
Resulting modules were then associated with four main CAD phenotypes as described  \cite{Hwang2005}: (1) extent of coronary atherosclerosis, (2) plasma levels of total cholesterol, LDL, VLDL, or HDL, (3) plasma glucose, HbA1c, insulin, or proinsulin levels, and (4) C-reactive protein levels which is an inflammatory marker. Next, CAD causality of phenotype-associated modules was examined by the inherited risk enrichment analysis using a total of six GWAS datasets: (1) CAD  \cite{Schunkert2011,Deloukas2013}, (2) fasting glucose  \cite{Dupuis2010}, (3) HbA1c  \cite{Soranzo2010}, (4) fasting proinsulin  \cite{Strawbridge2011}, (5) blood lipids  \cite{Teslovich2010}, and (6) type 2 diabetes  \cite{Burton2007}. For modules without sufficient number of cis-eQTLs ($n<10$), we instead used SNPs located in the vicinity of each module gene (+/-500kb of TSS) and adopted a new gene set enrichment analysis (GSEA)–like method  \cite{Subramanian2005}. CAD association \textit{P} values for each eQTL/SNP/module were assigned based on the corresponding SNP \textit{P} value in GWAS. A competitive null hypothesis was created from equally sized random sets of genes (n=1000) to enable comparing degree of CAD association in the STAGE-phenotype associated modules to the null distribution. The pipeline is summarized in Figure \ref{fig:crosstissScheme}.}
\subsection{A Cross-tissue Cross-species Validation of Regulatory Gene Networks for CAD Causality}
{The WGCNA revealed 171 modules; 94 were tissue specific, and 77 were cross-tissue modules. Of the 171 modules, 61 were associated with at least one of the four main CAD phenotypes in the STAGE patients. By analyzing these modules for inherited risk enrichment, GSEA-like score, and GWAS genes, we identified 30 CAD-casual modules. For example,  we found eight RGNs that were related to the extent of coronary atherosclerosis (assessed in pre-operative angiograms). Two of these eight RGNs found in SF had 10 and 15 CAD candidate genes identified by GWAS, respectively. Three of these 8 RGNs were found in AAW and enriched in various functional processes of high relevance for atherosclerosis (e.g., immune system processes, HDL cholesterol levels, and RNA-processing genes). These three RGNs were also and highly enriched in CAD association according to risk-enrichment analysis.}
\newline \indent
{In cross-species analysis, mouse orthologs to genes in 12 of the 30 CAD-causal modules were also associated with corresponding mouse phenotypes in the HMDP study  \cite{Bennett2010}. Three of these cross-species validated CAD-casual modules (RGNs 42, 58, and 98) were associated with the extent of both human and mouse atherosclerosis. Key drivers in these three RGNs were targeted in the THP1 foam cell model using siRNA where key drivers in RGN 42 was found to affect foam cell formation \textit{in vitro}.
}

\section{Paper IV: DNA Variants in Three Eicosanoid pathway genes}
{Three genes (ALOX5, ALOX5AP, and LTA4H) in eicosanoid pathway are believed to be involved in atherosclerosis and CAD. Expression of these genes is elevated in atherosclerotic plaques (both in human and mice) \cite{Cipollone2005,Mehrabian2002} , and SNPs in these genes are associated with carotid intima-media thickness  \cite{Dwyer2004}, myocardial infarction, CAD, and ischemic stroke  \cite{Helgadottir2004,Allayee2008,Helgadottir2006}. Despite their involvement in CAD/MI, no significant variants have been reported in these genes by GWAS . In this paper, SNPs in these three genes were extensively studied using two independent datasets: STAGE  \cite{Hagg2009} and the Athero-Express Biobank Study (AE)  \cite{Verhoeven2004}. Genotype imputation and cis-eQTL analysis methods (such as kruX) were used to analyze the SNP-gene associations. In sum, we found only one significant eQTL, in LTA4H in WB in the STAGE study. No significant eQTLs in these genes were identified in the AE dataset.}
\subsection{Identifying DNA variants in ALOX5, ALOX5AP, and LTA4H}
{STAGE and AE genotypes were imputed by using 1000 Genomes EUR population panel, and  SHAPEIT2  \cite{Howie2012, Delaneau2011} and IMPUTE2  \cite{Howie2012} softwares. In STAGE, kruX method was used for cis-eQTL inference as descibed.  SNPs within +/- 50kb of the three genes with at least 97\% call rate or imputation quality of > 0.9 were investigated.}
\subsection{There is no Association between DNA Variants and Three Eicosanoid Gene Expression}
{1,454 SNPs in AE and 1,078 SNPs in STAGE were examined. Consecutively, all the associations were corrected for multiple testing: for the plaque phenotypes: p = 0.05/(7 histological phenotypes $\times$ 1453 independent variants) = $4.92 \times 10^{−6}$, for serum protein level association with SNP: p = 0.05/(2 serum proteins $\times$ 1453 independent variants) = $1.72 \times 10^{−5}$ ; and finally for eQTL analysis: p = 0.05/(3 genes $\times$ 7 tissue types $\times$ 1078 independent variants) = $2.21 \times 10^{−6}$ pvalue cutoffs were used. In STAGE, only one eQTL (rs6538697) associated with LT4AH expression in WB  passed the multiple testing correction.}
\newline \indent
{In AE, two SNPs, rs4627178 for ALOX5AP and rs752059 for ALOX5, were associated with serum protein levels. Furthermore, rs9743326 was found associated with intraplaque microvessels, and rs17216508 with the number of smooth muscle cells in the atherosclerotic plaque as assessed in AE. However, none of these associations remained significant after correction for multiple testing.}
\begin{center}
{~}
\newline
\decoone
\end{center}

\chapter{Discussion}
{DNA variations identified by GWAS explain only a small proportion of the risk for CCDs  \cite{Bjorkegren2015,Schadt2012}. The possible reasons were discussed in Introduction (interactions between environmental and genetic factors, rare DNA variants, etc). New approaches, such as GGES, seek to utilize mRNA abundance as a sensor for DNA variation and thereby better understand how genetic variance affects clinical phenotypes as RNA abundance reflects the combined effects of DNA variation and tissue microenvironments. One of the core goals of GGES is to identify genome-wide eQTLs, which computationally is not an entirely trivial task. This is because there are approximately 8 million SNPs at the 5\% MAF cutoff  \cite{DbSNP}, and approximately ~20,000 protein-coding genes. If all possible associations are tested (i.e., all possible SNP-gene associations), $160 \times 10^{12}$ individual tests needed. One way to reduce the computational burden is to use matrix multiplications. This is the basis for eQTL mapping techniques, such as Matrix-eQTL  \cite{Shabalin2012}, to accelerate testing of billions of associations. The software tool developed in Paper I, kruX, aimed to address some shortcomings of Matrix-eQTL in detecting nonlinear associations (due to skewed genotype groups) and to find a method to reduce the sensitivity to data outliers. kruX is based on a nonparametric statistical test, Kruskal-Wallis, and uses matrix multiplications to speed up the computations. We found that the Kruskal-Wallis test used in kruX is more conservative and less sensitive to data outliers than the two models (ANOVA and additive linear models) used in Matrix-eQTL.}
\newline \indent
{In paper II, we used kruX to identify eQTLs in the STAGE dataset (consisting of seven CAD metabolic and vascular tissues) and inherited risk enrichment analysis to examine CAD-associated risk. We discovered that shared-tissue cis-eQTLs (those present in more than one tissue) were generally more enriched in CAD risk than tissue-specific eQTLs. This effect was even more pronounced for eQTLs active across many tissues (i.e., $>$ 4 tissues). This suggests that there is a complex interaction of genetic regulation of genes in CAD between metabolic and vascular tissues. We investigated the possibility that this higher enrichment for eQTLs found in multiple tissues is not a result of stronger association with the expression traits for multi-tissue eQTLs than tissue-specific. In testing this, we found that eQTLs with high \textit{P} values were not further enriched than eQTLs with low \textit{P} values regardless of the eQTLs were tissue-shared or tissue-specific. Moreover, there were no significant correlations between eQTL \textit{P} values (tested at FDRs 10\%, 15\%, and 20\% and MAF of 0.05, 0.10, and 0.15), and GWAS \textit{P} values. In sum, these results strongly support the notion that eQTLs active across multiple vascular and metabolic tissues are of stronger relevance for CAD risk and etiology than tissue-specific eQTLs.}
\newline \indent
{A comprehensive method, gene set enrichment analysis (GSEA), was proposed in 2005 to analytically score gene sets of interest based on common biological functions  \cite{Subramanian2005}. Since then, numerous methods and applications for such analysis were developed. SNP-based set enrichment analysis (SSEA), built on GSEA, was developed to study SNP sets instead of genes  \cite{Weng2011}. SSEA uses SNP information from GWAS to study enrichment of gene sets. In SNP-based approaches, SNPs for each gene are identified (e.g., distance to gene), a single \textit{P} value for each gene is assigned, and finally a score for gene sets is assigned through empirical permutation techniques or statistical tests.}
\newline \indent
{The possible problem with GSEA and SSEA techniques surfaces when SNPs are chosen solely on the basis of their physical distance from their gene. This distance can range from 5 kb to 500 kb in different publications. Inherited risk enrichment analysis uses positive features of SSEA and GSEA and integrates them into a single pipeline. In contrast to GSEA-like methods, inherited risk enrichment analysis uses eQTL information to assign SNPs to genes. eQTL-gene association is a more reasonable choice than assigning SNPs to gene based on a mere distance. Another issue with GSEA-based methods is gene set size. It is a potential source of bias, as the number of genes increases in a gene set, the possibility of higher risk scores increases. Some of the new approaches to GSEA try to correct for gene set size. In the inherited risk enrichment analysis pipeline that we developed, only eQTLs are taken into account, and any potential effect from gene set size is mitigated through permutations and pruning steps. Indeed, in Paper II, we show that gene sets with a large number of genes were generally not more enriched than gene sets with a small number of genes.}
\newline \indent
{With GSEA and GSEA-like methods, depending on the gene length and gene set size, SNPs that are assigned to each gene might be redundant. For example, if genes are closer and the search windows are overlapping, same signal (i.e., SNP \textit{P} value) from overlapping regions will be selected multiple times. Thus, inflation of results due to redundant SNPs can be an issue. Our inherited risk enrichment analysis operates only at the SNP level, and mitigates the redundant SNPs issue by LD pruning after eQTL expansion. (SSEA implements an LD-pruning step, but it still doesn’t consider overlapping genes.) Our approach also removes any potential bias from gene length and inflation or deflation of results due to LD between SNPs and their statistical significance in GWAS.}
\newline \indent
{The inherited risk enrichment analysis method also includes permutation and resampling approaches. These approaches can be implemented at three levels: genes, individuals, and SNPs. At the gene level, indexes of genes are randomized and resampled. This is suitable only when analyzing gene sets at the gene level, and it doesn’t consider SNP distribution nor LD structures between SNPs. At the individual level, permutation is computationally intensive and requires access to individual level data. In this approach, indexes for the individuals are permuted and test statistics for each SNP are re-calculated, then for each gene set a new test statistics is assigned. Since individual level data is not always accessible and is computationally intensive it is rarely used. The SNP level approach, however, is more similar to gene level permutation. Either SNP indexes are permuted or they are resampled based on certain criteria. It is easier to maintain LD structure between SNPs, MAF, and chromosomal distribution of the SNPs, and number of the SNPs when permuting or resampling at the SNP level. In our inherited risk enrichment analysis, we therefore used resampling at the SNP index level.}
\newline \indent
{A possible drawback of inherited risk enrichment analysis is its reliance on eQTL information from the STAGE study. Almost two thirds of genes in STAGE dataset didn’t have any eQTLs. Thus, if there are fewer than about 10 eQTLs/gene set, this method is not useful. The lack of sufficient number of eQTLs/SNPs to analyze gene sets is, however, also a challenge for GSEA and SSEA methods. A larger eQTL datasets become available the problem with lacking eQTLs for genes will be reduced.}
\newline \indent
{In sum, the inherited risk enrichment analysis provides a powerful tool integrating  GGES and GWAS to assign disease relevance to lists of genes as judging from their  combined association with disease. In this thesis we used the STAGE GGES and GWAS to implement this strategy for CAD. However, as GGES will be performed on other complex traits and CCDs, this new data-driven unbiased approach of extracting disease relevance for lists of genes/transcripts (e.g., defined by differential expression on up to gene networks), we think, has the potential to grow in importance and utility. Inherited risk enrichment analysis (and similar methods that use eQTL data) is also a powerful approach to study disease causality. In paper III, gene modules were considered causal if any of the genes in the module were known CAD candidate genes for genome-wide loci identified by GWAS or if the module eQTLs were enriched in CAD risk according to our inherited risk enrichment analysis reusing GWAS datasets. The notion that disease association also refers causality is principally based on the central dogma in biology. In brief, the central dogma states that genetic variation always is upstream of (e.g. causal for) gene expression that in turn is upstream of variation in the phenotype (e.g., disease). As a consequence, SNPs are always causal for gene expression and from that it can be inferred that genes with eQTLs are more likely to also to affect the disease (i.e., to be causal) and not to be regulated by disease (i.e., reactive). Adding the GWAS disease information, we can further refine this logic to specifically seek causality for different diseases (e.g., CAD) or phenotypes (e.g. LDL levels).}
\newline \indent
{In Papers II and III, the combination of the STAGE study and inherited risk enrichment analysis was used to identify new CAD target genes and gene networks. The unique multi-tissue STAGE dataset can also be used to further validate the importance of more established atherosclerosis and CAD candidate genes. This was one of the main goals in Paper IV. Specifically, the STAGE dataset was used to study three previously known genes (ALOX5, ALOX5AP, and LTA4H) involved in the inflammation-mediating eicosanoid pathway associated with atherosclerosis development. By analyzing the effect of common variants on gene expression (i.e., eQTLs), circulating protein levels, and several atherosclerotic plaque phenotypes, we found only one significant association between these parameters in the form of a eQTL in whole blood for LTA4H. These results are in line with GWAS of  CAD/MI \cite{Schunkert2011} where these genes have not been reported. Thus, paper IV is a first indication that eQTL mapping and consequently using these eQTLs in integrative analysis re-using GWAS datasets in the form of inherited risk enrichment analysis, is not only a data-driven (i.e., unbiased) approach to identify novel candidate genes and gene networks underlying complex traits like CAD but also tools to re-assess the importance of established disease risk genes. Importantly, however, the fact that a gene is not genetically regulated (e.g., have eQTLs) that also are associated with a complex trait according to GWAS, does not rule out a causative role in disease/phenotype development. In fact, we generally observe that key drivers of gene networks associated with CAD (Paper III) are not in themselves regulated by eQTLs associated with CAD. A plausible explanation for this observation might be negative selection of genetic variants affecting key regulatory genes through evolution as such variation may not be in agreement with successful breeding (e.g. cause embryonic or early life lethality).}
\begin{center}
{~}
\decoone
\end{center}

\chapter{Concluding Remarks}
{In this thesis, we have integrated the unique GGES STAGE with GWAS using a system biological approach  \cite{Ideker2001} to increase our understanding of CAD. We think this approach is powerful as it does not consider single gene associations with CCDs, which risk rendering spurious correlations, but instead utilizes data-driven, whole-genome and functional approaches that are needed for a better global understanding of the complex inheritance and etiology of CCDs. Specifically, we think the eQTL mapping technique; kruX and the inherited risk enrichment analysis methods proposed in this thesis should be useful to improve our understanding of other CCDs besides CAD.}
\newline \indent
{In the future, similarly as we historically have been integrating single DNA variants, genes, proteins and/or metabolites with disease phenotypes, we foresee that CCD understanding will be faster advanced by also integrating these features a the genomic, transcriptomic, proteomic, metabolomic levels with phenotypic/clinical data. To do so we do not only need more cost-effective and reliable screening methods (e.g., RNA sequencing) but also better and more robust computational tools to analyze and integrate these huge datasets. In the end, as in the case of individual genes, functional gene sets, as gene networks with key drivers, need further validation both by using in silico experiments to model the dynamics of the interaction between components in these networks and in the end, by using efficient means of in vitro and in vivo validation experiments.}
\newline \indent
{The four studies presented in this thesis each cover some aspects of systems biological steps needed to better understand CCD inheritance and etiology. However, they are also limited to the use of eQTL datasets alone. Thus, falls short when it comes to analyzing other modes of risk inheritance such as epigenetics and non-RNA intermediate phenotypes such as protein abundance and metabolites. As the final aim needs to analyze and map the entire variation and sources of risk inheritance and disease etiology, epigenetics and non-RNA intermediate phenotypes need also to be considered. Emerging improvements in high-throughput techniques to analyze protein abundance  \cite{Zhang2014,Vogel2012} and advances in epigenetics \cite{Romanoski2015,Kundaje2015,Bernstein2010} will pave the way for also integrate these sources of variance in systems biological approaches to CCDs.}
\begin{center}
\decoone
\end{center}

\chapter{Acknowledgement}
{
\begin{flushright}
\textit{``I conclude with a warm 'hello' to those friends \\who don't care about this book,\\ who won't read the book, \\and who had to put up with me while I wrote it" \\
Alan Chalmers}
\end{flushright}
{There have been many people during the past four years, Friends, colleagues, and acquaintances who influenced me both in my personal life and in my research. I want to thank and acknowledge all of them for making science and life more enjoyable. I would like to extend my most sincere gratitude towards my supervisor {Johan Bj\"{o}rkegren}; a great scientist and mentor. I always admired your consistency, planning, decisiveness, and leadership. I think science truly needs your ideas. The way you deal with research problems, with an open mind to explore new solutions, will always be my inspiration. It was an honor to take a small part in your research for five years, it was a short but lovely years of my life. Thank you for giving me the opportunity and trusting in me.}
{My co-supervisor {Josefin Skogsberg}, a great scientist and researcher. I've learned a lot from you. It's too bad you're leaving academia but I wish you a successful career wherever you go, which I'm sure you will be.}\\
{My co-supervisor {Tom Michoel}, a master problem solver and a great computational biologist. Whenever I was stuck on a project, the solution was just an email away and you were there to rescue. When I met you, I understood why they always said smart physicists go to biology. It was a pleasure to work under your supervision, and I hope to get a chance to work with you again in the future.}\\
{My co-supervisor {Christer Betsholtz}, for whom I have the utmost respect. Unfortunately, I didn't have the chance to directly work with you, and I regret that, but I hope to get the opportunity to do so in the coming years. Also I would like to specially thank {Ulf Eriksson}, my external mentor, for being there when I needed it.}\\
{A thesis reflects as much scientific growth, as it reflects personal growth. There were many people who were there to make this thesis possible, from casual lunchtime chats to discussion on science and life. It was not possible for me to get to know everyone around myself, let alone acknowledging every single one of them. Here I decide to mention a minimal collection of people. First of all, I'd like to name my colleagues and office mates {Husain Talukdar} and {Ar\'{a}nzazu Rossignoli}, for their friendship and companionship. Secondly, people from other labs in MBB (in no particular order): {Chad Tunell}, {Sebastian Lewandowski}, {Christine M\"{o}ssinger}, {Linda Fredriksson}, {Lars Jakobsson}, {Daniel Nyqvist}, {Azadeh Nilchian}, {Jong Wook Hong}, {Chenfei Ning}, {Tian Li}, {Mirela Balan}, {Joanna Wisniewska}, {Agnieszka Martowicz}, {Patricia Rodriguez}, {Jing Guo}, {Sonia Zambrano}, {Olle Rengby}, {Amirata Saei Dibavar}, {Bo Zhang}, {Pierre Sabatier}, {Mohammad Pirmoradian}, {Xiaoyuan Ren}, {Goncalo Castelo-Branco}, {Katja Petzold}, {Alfredo Gimenez-Cassina}, and {Ahmad Moshref} and many more including all {Vascular Biology}.}\\
{During past years, there were people that their friendship extended beyond science and research. These amazing friends, in no specific order, are: {Roman Valls}, {Spyridon Gkotzis}, {Arnold Konnerth}, {Mattias Fr{\aa}nberg}, {Farzad Abtahi}, {Sahar Khanmohamadi}, {Mahya Dezfouli}, {Fatemeh Ostad Ebrahim}, {Milad Razzaghpour}, {Yasi Golabkesh}, {Behrouz Afshari}, {Serveh Sadeghi}, and {Amir Mirzaei}. We shared so many good memories during past years, thank you for your friendship.}\\
{I hope I will not upset anyone by singling out {Amarendra Badugu}, a true friend and an amazing person to talk to. Thank you for helping me to keep my chin up when I was feeling down.}\\
{My lovely wife, {Afsaneh}, How can I be thankful for all your support, love, and company? It would have been impossible to do it all without you by my side. My amazing sisters: {Fateme} and {Zahra}. They say, there is no better friend than a sister, and I have two best friends. You are the best. And {Ali Nagilu}, you are the brother I always wanted to have. Finally, my {parents} for giving me the freedom I need to explore science, and teaching me to think critically. Thank you for your endless love and support.}
\begin{center}
\decoone
\end{center}
\begin{center}
{
\centering
\vspace{3cm}
{Hassan, May 2016, Stockholm}\\
\vspace{5cm}
{This thesis was typeset using \LaTeX}
}
\end{center}
}
\bibliographystyle{unsrtetal}
\bibliography{thesis}
\begin{center}
\decoone
\end{center}


\end{document}